\documentclass[12pt]{article}
 \usepackage[latin1]{inputenc}
 \usepackage{verbatim}
\usepackage{amssymb}
\usepackage{amsmath}
\usepackage{amsfonts}
\usepackage{bm}
 \usepackage[dvips]{color}
\newtheorem{theorem}{Theorem}[section]
\newtheorem{proposition}{Proposition}[section]
\newtheorem{lemma}{Lemma}[section]
\newtheorem{corollary}{Corollary}[section]
\newtheorem{remark}{Remark}[section]
\newtheorem{assumption}{Assumption}[section]

\newcommand{\proof}{{{\bf Proof}}}
\newcommand{\qedrem}{\hfill\framebox[0.2cm]}
\newcommand{\qed}{\hfill\rule{2mm}{2mm}}

\newcommand{\ze}{\epsilon}

\newcommand{\p}{\partial}
\newcommand{\pt}{\partial}

\newcommand{\calc}{{\cal C}}
\newcommand{\calp}{{\cal P}}
\newcommand{\calq}{{\cal Q}}

\newcommand{\N}{{\ensuremath{\mathrm{I}\!\mathrm{N}}}}
\newcommand{\R}{{\ensuremath{\mathrm{I}\!\mathrm{R}}}}

\renewcommand{\O}{{\Omega\,}}

\newcommand{\zf}{\varphi}

\newcommand{\zt}{\tau}
\newcommand{\baf}{\bar\zf}
\newcommand{\fibar}{\bar \varphi}
\newcommand{\psibar}{\bar \psi}
\newcommand{\fitil}{\tilde \varphi}
\newcommand{\psitil}{\tilde \psi}

\newcommand{\qacq}{{\cal Q}({\cal A+C}){\cal Q}}
\newcommand{\qsq}{{\cal Q}S{\cal Q}}
\newcommand{\qsp}{{\cal Q}S{\cal P}}
\newcommand{\psp}{{\cal P}S{\cal P}}
\newcommand{\psq}{{\cal P}S{\cal Q}}

\newcommand{\Mbar}{\overline M}
\title{\bf 
Rigorous drift-diffusion asymptotics of a high-field quantum transport equation}
\author{\\[-0.4cm]
{\bf Chiara  Manzini} and {\bf Giovanni Frosali} \\
Dipartimento di Matematica ``G.Sansone" \\
Universit\`a di Firenze - Via S.Marta 3\\
I-50139 Firenze, Italy\\
}
\date{}
\begin{document}
\maketitle
%
%

\begin{abstract}
The asymptotic analysis of a linear high-field Wigner-BGK equation is developped by a modified Chapman-Enskog procedure.
By an expansion of the unknown Wigner function in powers of the Knudsen number $\epsilon$, evolution equations are derived for the
terms of zeroth and first order in $\epsilon$. In particular, it is obtained a quantum drift-diffusion equation for the position density,
which is corrected by field-dependent terms of order $\epsilon$.  Well-posedness and regularity of the approximate problems are established,
and it is proved that the difference between exact and asymptotic solutions is of order $\epsilon ^2$, uniformly in time and  for arbitrary
initial data.\\
\mbox{}
\end{abstract}

\section{Introduction}
\setcounter{equation}{0}
Quantum mechanics has recently proved 
an essential tool for modeling 
the new generation of nanodevices 
\cite{markringhsch}.
However, the adoption of quantum models 
requires a delicate compromise 
with quantum statistics principles. Hamiltonian dynamics is described at the quantum level, either in terms of
wave-functions (via Schr\"odinger-Poisson-systems), or of density-matrix operators (via von Neumann equation).
For different reasons, both formulations are not suitable for simulations:
precisely, the wave-function approach can not be extended to picture dissipative dynamics of open quantum systems, while the density matrix approach is not appropriate to describe finite position domains, due to its non-local character.
For the same reasons, it is instead convenient to employ (Wigner) quasi-distribution functions 
\cite{arnoldjuengel,Wig32}.
Nevertheless, a phase-space description of a multi-dimensional dynamics presents well-known computational drawbacks. On the other hand, quantum hydrodynamic models
seem to be a promising tool both from the numerical and the analytical point of view \cite{juengel,juengelpinnau1}. Similarly, in semi-classical semiconductor theory, the interest of modelists has shifted from
Boltzmann equation to hydrodynamic systems, and they have been widely studied both for a physical validation and from an analytical and numerical point of view (cf.~\cite{anileromano} and the references therein).
A rigorous derivation of quantum hydrodynamic models from more fundamental ones, in either Schr\"odinger or Wigner formulation, is an open and analytically demanding problem
\cite{DeMeRi05,gasmar,JueMa}. This is the motivation of the present paper.\\ 
The preliminary step for passing from the kinetic picture to a macroscopic one consists in including dissipative mechanisms in the
evolution model, for example, 
the interaction of the quantum system with the environment. In the weak coupling limit, 
a Markovian dynamics can still be adopted, and the description via an (operatorial) evolution equation in Lindblad form is considered quantum-physically correct \cite{Lindblad}.
From this class of evolution equations, kinetic models of open systems can be derived via Wigner transform.
In Section 2 we shall briefly review the most popular Wigner models of irreversible dynamics.\\
In this paper, we consider the case of an open quantum system in a high-field regime, more precisely, of
an electron ensemble subject to an external potential, whose effect is comparable with the interaction with the ion crystal.
Including high-field effects has great relevance in semiconductor simulation. A macroscopic model of this evolution is expected to contain
field-dependent transport parameters, that are tipically deduced via fitting procedures.
We refer the reader to \cite{{DeJu}} for an updated review of derivations
of {semi-classical} high-field drift-diffusion models by diverse limit procedures: in particular, in \cite{DeJu}, from an energy-transport
model, are obtained explicit field-dependent mobilities.
On the contrary, in \cite{benab},  a high-field drift-diffusion model with non-explicit field-dependent coefficients is derived, as the limit of a Spherical Harmonics Expansion of semi-classical Boltzmann equation.\\ 
We present a {rigorous} derivation 
of a Quantum Drift-Diffusion (QDD) equation with {explicit} field-dependent mobility and diffusion coefficient.
We shall start from the Wigner equation with an additional linear BGK term, modeling the interaction with the environment, and then adapt the equation to the high-field case, by rescaling it in terms of the Knudsen number $\ze$.
Thus, our contribution is the quantum counterpart of \cite{poupaud}.
We recall that, in \cite{GardRi}, the starting point is the Wigner-BGK equation as well, but collisions are considered to be the strongest
mechanism during the evolution (moderately high-field regime), and the relaxation term is derived via a Chapman-Enskog procedure.
In our case, the additional relaxation term is instead an $\mathcal{O}(\hbar^2)-$approximation of the Wigner-transformed relaxation term
in operatorial formulation (cf.~Section 2).
Moreover, we perform an asymptotic expansion of the unknown Wigner function in terms of $\ze$, according to a {modified} Chapman-Enskog
procedure introduced in \cite{banasiakmika95}.
This method has been applied to many kinetic models and constitutes a valuable tool for a {rigorous} asymptotic derivation of macroscopic
models (cf.~Section 5).
We substitute the Wigner unknown in the
originary evolution problem, with the expansion of order $\ze^2$, and we get an approximated problem: in particular, an equation with unknown
the electron position-density.
This equation is precisely the QDD equation corrected by the $\mathcal{O}(\hbar^2)$-Bohmian term of order $\ze$, and by field-dependent terms, of order $\ze$ as well. These terms contain the same field-dependent coefficients obtained in the semi-classical case \cite{DeJu,poupaud}.\\
The well-posedness of the $\mathcal{O}(\ze^2)$-approximated problem is discussed in Sections 7 and 8, and finally, in Section 9,
we prove that the difference between the solutions of the originary and of the approximated evolution problems is also of order $\ze^2$.
In conclusion, with the present analysis we obtain a QDD equation with field-dependent mobility and diffusion coefficients and we prove
rigorously that, up to a certain degree of accuracy, it constitutes a model of quantum transport in the high-field case.
From the analytical point of view, this equation is a second-order parabolic PDE with { non-homogeneous} coefficients.
In particular, it belongs to the class of singularly perturbed equations; accordingly,
the well-posedness result, together with the regularity estimates derived in Section 8, are complementary to the discussion
in \cite{banasiakAAM} about the same class of equations with constant coefficients. 
A counterpart of our analysis is the well-posedness study of the quantum drift-diffusion equation,
in the fourth-order formulation obtained via a ``classical-equilibrium'' approximation \cite{juengelpinnau1}.
We remark that the asymptotic procedure used here presents analogies with the Chapman-Enskog one in kinetic theory; nevertheless,
it is well-known that the latter does not deal with the ``initial layer'' problem, namely, the instants close to the initial one are
excluded from the analysis, due to the rapid changes of the solution \cite{poupaud}. In the present approach instead, 
the initial layer problem is solved at once. 

\section{Wigner-BGK equations}
\setcounter{equation}{0}

Let us consider a quantum system with $d$ degrees of freedom, evolving
under the effect of an external potential $V=V(x), \,x\in \R^d$.
The Wigner equation with unknown the quasi-distribution function $w=w(x,v,t), (x,v)\in \R^{2d}, \, t>0 $,
provides a kinetic description of the evolution of the system.
It reads
\begin{equation}
\label{eq:Wig}
\frac{\partial w}{\partial{t}} + v\cdot\nabla_xw - \Theta[V]w \;\: = \;\:
0,\quad (x,v)\in \R^{2d}, \quad t>0 \,,
\end{equation}
with
the pseudo-differential operator $\Theta[V]$ defined by
\begin{eqnarray}
\label{eq:theta} (\Theta[V]w)(x,v,t) & = & \frac{i}{(2\pi)^d} \int_{\R^d}\!\int_{\R^d}\delta
V(x,\eta)w(x,v^{\prime},t)e^{i(v-v^{\prime})\cdot\eta}\,dv^{\prime}\,d\eta \nonumber\\
 & = & \frac{i}{(2\pi)^{d/2}}\int_{\R^d}\delta V(x,\eta){\cal
F} w(x,\eta,t)e^{iv\cdot\eta}\,d\eta\,,
\end{eqnarray}
where
$$
\delta V(x,\eta) := \frac{1}{\hbar}\left[ V\left(x+\frac{\hbar\eta}{2{m}}\right) -
V\left(x-\frac{\hbar\eta}{2{m}}\right)\right]
$$
and ${\cal F} f (\eta)\equiv [{\cal F}_ {v\to \eta}f](\eta)$ denotes
the Fourier transform of $w$ from $v$ to $\eta$.
In the Fourier-transformed space $\R^d_{x}\times\R^d_{\eta}$ the operator $\Theta[V]$ is the multiplication
operator by the function $i\,\delta V$; in symbols,
\begin{equation}
\label{eq:product}
{\cal F}\left(\Theta[V]w\right)(x,\eta) = i\,\delta V(x,\eta){\cal F} w(x,\eta)\,.
\end{equation}
Eq.~\eqref{eq:Wig} corresponds via Wigner-transform to the von Neumann equation describing the conservative dynamics of an {isolated} quantum system \cite{Wig32}.
Successive modifications of the Wigner model have been proposed to picture an irreversible interaction of the system with the environment. In \cite{FroMaRi} a scattering term is derived  by a weak-coupling limit; however, due to its non-locality, 
it is not suitable for simulations and for mathematical analysis.
A second possibility is an additional diffusive
term, as in the quantum counterpart of Fokker-Planck (FP) equation of classical kinetic theory \cite{CaLe} (cf.~\cite{CaErFro} for the latest derivation and \cite{ADM06,ADM06b} for
the latest well-posedness results). Unlike the Wigner equation with the scattering term, the quantum FP equation is the Wigner-transformed version of a Markovian master equation in Lindblad form, namely, it is the kinetic version of a quantum-physically correct model \cite{arnoldsparber}. 
The shape of the drift-diffusion equations corresponding to the low-field, respectively high-field, scaling of the classical, respectively quantum, FP equations are presented in \cite{arnold_limit}.
Another possibility is to insert a BGK operator, either linear or non-linear, like in \cite{bonilla}, meaning that after a time $1/\nu$ the
system relaxes to a prescribed state $w_{\mathrm{eq}}$; namely,
\begin{equation}
\label{eq:WigBGK}
\frac{\partial w}{\partial{t}} + v\cdot\nabla_xw - \Theta[V]w \;\: = \;\:
- \nu(w-w_{\mathrm{eq}})\,,\quad (x,v)\in \R^{2d}, \quad t>0.
\end{equation}
In the recent literature \cite{DeMeRi05,Gard94, GardRi, JueMa}, diverse relaxation-time states $w_{\mathrm{eq}}$ have been proposed.\\
The standard picture is that the system converges to a state of thermodynamical equilibrium with the surrounding environment at temperature $T$. The operator that individuates the statistical equilibrium state at (constant) temperature $T=1/k\beta\,$ ($k$ is the Boltzmann constant) is
$
\mathrm{e}^{-\beta H},
$
$H$ being the energy operator associated to the system. The von Neumann equation modified by a relaxation-time term containing $\mathrm{e}^{-\beta H}$
is {in Lindblad form} \cite{arnoldrelax}. Accordingly, a Wigner-BGK model, being
the Wigner-transformed version of that equation (i.e., containing the Wigner-transformed of $\mathrm{e}^{-\beta H}$ as relaxation-time state), formally belongs to the class of quantum-physically correct kinetic models.
In his pioneer article \cite{Wig32}, E.~Wigner
applies an expansion in terms of $\hbar$ to the Wigner function corresponding to the operator $\mathrm{e}^{-\beta H}$, and obtains the classical equilibrium distribution function on the phase space with correction of non-odd order in $\hbar$:
\begin{multline}
\label{eq:wig_eq}
w_{\mathrm{W}}(x,v):=\left(\frac{m}{2\pi\hbar }\right)^{d}\!\! e^{-\beta {\cal E}} \\
\times\left\{1+{\hbar^2}\frac{\beta^2}{24}\left[-\frac{3}{m} \Delta V+\frac{\beta}{{m}}|\nabla V|^2 +
\beta \sum_{r,s=1}^dv_rv_s\frac{\partial^2 V}{\partial x_r\partial x_s}\right]+{\cal O}(\hbar^4)\right\}\,,
\end{multline}
where ${\cal E}(x,v):= mv^2/2+V(x)$ is the total energy of the system.
Let us call $w_{\mathrm{eq}}$ its {local (in time and space)} version, defined by
$$
w_{\mathrm{eq}}(x,v,t)\:\;:=\:\; C(x,t)\ w_{\mathrm{W}}(x,v)\,,
$$
with $C$ to be chosen. By assuming
\begin{equation}
\label{eq:constraint}
\int w_{\mathrm{eq}}(x,v,t)\,dv \:\;=\:\; \int w(x,v,t)\,dv  \:\;=:\:\;  n[w](x,t)\equiv n(x,t)\,,
\end{equation}
and since, by the direct computation,
$$
\int\! w_{\mathrm{W}}(x,v)\,dv=\left(\frac{m}{2\pi\hbar^2\beta}\right)^{d/2}\!\!\!\!{e^{-\beta V}}
\left\{1+\hbar^2\frac{\beta^2}{12{m}}\left[ - \Delta V+\frac{\beta}{2}|\nabla V|^2\right]+{\cal O}(\hbar^4)\right\},
$$
the local Wigner thermal equilibrium function $w_{\mathrm{eq}}$ equals
\begin{multline}
\label{eq:wig_eq_approx}
w_{\mathrm{eq}}(x,v,t)=n(x,t)\left(\frac{\beta m}{2\pi}\right)^{d/2} e^{-\beta mv^2/2}  \\
\times\left\{1+\hbar^2\frac{\beta^2}{24}\left[-\frac{1}{m}\Delta V+\beta \sum_{r,s=1}^{d}v_rv_s\frac{\partial^2 V}{\partial x_rx_s}\right]+{\cal O}(\hbar^4)\right\}\,.
\end{multline}
In \eqref{eq:wig_eq_approx} can be recognized the classical (normalized) Maxwellian 
\begin{equation}
\label{eq:Maxw}
F(v):=\left(\frac{\beta m}{2\pi}\right)^{d/2}\!\!{e^{-{\beta m v^2}/{2}}}\,,
\end{equation}
parametrized by the density $n$ and the constant temperature $1/k \beta$, with an additional correction term of order $\hbar^2$. We shall consider the expression \eqref{eq:wig_eq_approx} as the ${\cal O}(\hbar^2)$-ap\-proxi\-ma\-tion of the Wigner function associated to the state to which the quantum system shall 
approach.\\

In \cite{DeRi} is presented an alternative strategy to individuate the relaxation-time state, that is the extension to the quantum case of Levermore's one for classical kinetic equations (\cite{Levermore}, cf.~\cite{anileromano}
for semi-classical equations). It consists in tackling a constrained minimization problem for the relative entropy of the quantum system under consideration, with respect to the environment. In the quantum case the procedure is performed at the 
operatorial level, due to the non-local definition of the entropy, in terms of the operators describing the states of the quantum system. However, the constraints for the minimization procedure are considered at the kinetic level. Thus, the Wigner transform ${\cal W}$ is used intensively to pass from the operatorial formulation to the kinetic one, any time it is required by the procedure. Due to that, the expression of the minimizer of the entropy formally derived in \cite{DeMeRi05} is non-explicit. Nevertheless, in \cite{DeMeRi05}, is formally proved that ${\cal W}\{\exp {\cal W}^{-1}f\}=\exp f + {\cal O}(\hbar^2)$ with $f$ defined on the phase-space. Accordingly, the (formal) minimizer reads
\begin{multline}
\label{eq:wig_deg}
w_A(x,v,t)\:\;:=\:\; e^{(A-\beta mv^2/2)} \\
\times \left\{1+{\hbar^2}\frac{\beta^2}{8}\left[+\frac{1}{m} \Delta A+\frac{\beta}{3{m}}|\nabla A|^2
+\frac{\beta}{3}\sum_{r,s=1}^dv_rv_s\frac{\partial^2 A}{\partial x_r\partial x_s}\right]+{\cal O}(\hbar^4)\right\}
\end{multline}
with $A=A(x,t)$ Lagrange multiplier used for the constrained minimization procedure, i.e.
$$
\int w_{A}(x,v,t)\,dv \:\;=\:\; n(x,t)\,.
$$
By comparison of the expression \eqref{eq:wig_deg} with \eqref{eq:wig_eq_approx}, it can be easily seen that they coincide if one identifies the Lagrange multiplier $A$ with $-\beta V$. In \cite{JueMa} it is indeed proved that $A=-\beta V + {\cal O}(\hbar^2)$ holds.
\begin{remark}
\label{remark:L2}
\em
It is crucial to recall that the correspondence via Wigner-transform of the operatorial and the kinetic formulations is merely formal, unless certain assumptions are posed both on the Wigner functions and on the operators \cite{lions}. On this point depends the analytical difficulty in stating rigorously the well-posedness of the strategy of derivation in \cite{DeMeRi05}. For the same reason, the analysis of Wigner equations is set in the Hilbert space $L^2$, since the necessary condition for the rigorous correspondence is satisfied \cite{lions} (cf.~\cite{ADM06b}, e.g.). \qedrem\\
\end{remark}
As a consequence of the previous discussion, in the present article we shall adopt the Wigner-BGK equation \eqref{eq:WigBGK} containing \eqref{eq:wig_eq_approx} on the right-hand side as the model of the open quantum system evolution. In particular, we remark that we shall consider the operator on the right-hand side as an ${\cal O}(\hbar^2)$-approximation, in the kinetic framework, of the dissipative dynamics induced by the interaction with the environment.
\section{The high-field Wigner-BGK equation}
\label{formulation}
\setcounter{equation}{0}
Our aim is to describe an open quantum system subject to a strong external potential; in particular, the action of the potential is to be considered comparable with the interaction with the environment. In order to adapt the Wigner-BGK equation \eqref{eq:WigBGK} to this specific case, we rewrite it by using dimensionless variables and, for this purpose, we introduce the time-scales of the action of the external potential and of the interaction with the environment.
Let us call $t_V$ the potential characteristic time
and $t_C$ the mean free time between interactions of the system with the background.
Then we introduce $x'=x/x_0$, $v'=v/v_0$,
$t'=t/t_0$, with $x_0, v_0, t_0$ characteristic quantities, and we call $w'=w(x',v',t')$ the rescaled Wigner function
(observe that we can indeed neglect to rescale the Wigner function).
Thus, we obtain
$$
\frac{x_0}{v_0t_0}\frac{\partial}{\partial{t}}w + v\cdot\nabla_xw - \frac{x_0}{v_0 t_V}\Theta[V]w \;\:
= \;\:
- \frac{x_0}{v_0 t_C} \nu (w-w_{\mathrm{eq}})\,,\quad t>0,\quad (x,v)\in \R^{2d},
$$
where we have omitted the prime everywhere.
If we introduce the relation $x_0=v_0 t_0$, we obtain
$$
\frac{\partial}{\partial{t}}w + v\cdot\nabla_xw - \frac{t_0}{t_V}\Theta[V]w \;\: = \;\:
- \frac{t_0}{t_C}  \nu \left( w-w_{\mathrm{eq}} \right) \,,\quad t>0,\quad (x,v)\in \R^{2d}.
$$
In the following we assume that the times $t_V$ and $t_C$ are comparable, in the sense
\begin{equation}
\label{eq:scaling}
\frac{t_V}{t_0} \approx \frac{t_C}{t_0} \approx \epsilon \,,
\end{equation}
where $\epsilon:={l}/{x_0}$ is the Knudsen number, since $l:=v_0 t_C$ is the characteristic length corresponding to the classical mean free path. This corresponds to say that the external potential and the interactions coexist during the evolution. In particular, $\ze \approx 0$ corresponds to an evolution in which the effect of the interactions is dominant on the transport ($t_C <\!\!< t_0 $ or equivalently $l <\!\!< x_0$). However, at this time the action of the external potential has the same strength, due to the assumption \eqref{eq:scaling} ($t_V <\!\!< t_0 $). In fact, the resulting equation is
\begin{equation}
\label{eq:Wigeq_adim}
\ze \frac{\partial w}{\partial{t}} + \ze v\cdot\nabla_xw - \Theta[V]\,w \;\: = \;\:
- \nu \left(w-w_{\mathrm{eq}}\right) \,,\quad t>0,\quad (x,v)\in \R^{2d}.
\end{equation}
We recall that it is the quantum counterpart of the one studied by F.~Poupaud in \cite{poupaud}.\\
Now we put \eqref{eq:Wigeq_adim} in abstract form. As motivated in Remark \ref{remark:L2}, a suitable setting for problems in Wigner formulation
is the Hilbert space $L^2({\R}^{2d})$. However, in order to give a rigorous sense to the expression
\begin{equation}
\label{eq:density}
n(x)\:\;:=\:\;\int w(x,v)\,dv\,, \quad \forall x\in\R^d,
\end{equation}
which enters the equation via the definition \eqref{eq:wig_eq_approx} of $w_{\mathrm{eq}}$,
we introduce the subspace  $X_k:=L^2({\R}^{2d},(1+|v|^{2k})\,dx\,dv; \R)$, with $k\in \N$, endowed with the norm
$$
\|w\|_{X_k}^2 =
\int_{\R^{2d}}\!\!\!  |w(x,v)|^2 (1+|v|^{2k})\, dx \,dv\,.
$$
Let us call $X_k^v$ the Hilbert space $L^2({\R}^{d},(1+|v|^{2k})\, dv; \R)$ and $H_k^m$ the Sobolev space $H^m_x \otimes X_k^v$. The weight $k$ has to be chosen according to the space dimension $d$: we call $d$-admissible
$$
k\in \N \quad \hbox{such that}\; 2k>d\,.
$$
The definition \eqref{eq:density} is well-posed for all $w\in X_k$ with $d$-admissible $k$, since
$$
\left|\int_{\R^{d}}\!\!w(x,v)\, dv\right| \leq C(d,k) \left(\int_{\R^{d}}\!\!|w(x,v)|^2(1+|v|^k)^2\, dv\right)^{1/2}\!,\quad \forall\, x\in \R^d
$$
by H\"older inequality (cf.~\cite{MB}). We define the streaming operator $S$ by
$$
{S}w = -v \cdotp \nabla_x w \;,\;
D(S)=\left\{w \in { X}_k\,|\, S \,w \in { X}_k \right\}\,,
$$
and the operators 
\begin{equation}
\label{eq:abstractoperators}
{\mathcal A}w:=\Theta[V]w,\qquad\calc w :=- (\nu \, w-\O w)\,, \quad \forall\, w\in X_k
\end{equation}
with the operator $\Omega$ defined by
$$
\O w (x,v):= \nu F(v)
\left\{1+\hbar^2\frac{\beta^2}{24}\left[-\frac{1}{m}\Delta V
+{\beta}\sum_{r,s=1}^{d}v_rv_s\frac{\partial^2 V}{\partial x_rx_s}\right]\right\}
\int\!w (x,v^{\prime})\,dv^{\prime}\,.
$$
The function $F(v)$ is the normalized Maxwellian, given by \eqref{eq:Maxw}.
Observe that we substitute the function $w_{\mathrm{eq}}$ defined in~\eqref{eq:wig_eq_approx} with the operator
$\Omega w$, that differs from $w_{\mathrm{eq}}$ by terms of order $\hbar^4$. Let us call $F^{(2)}$ the $O(\hbar^2)-$coefficient in the above definition of $\Omega$ 
$$
F^{(2)}(x,v)\equiv F^{(2)}[V](x,v)=\frac{\beta^2}{24}\left[-\frac{1}{m}\Delta V
+{\beta}\sum_{r,s=1}^{d}v_rv_s\frac{\partial^2 V}{\partial x_rx_s}\right]F(v)\,,
$$
such that
$$
\Omega w (x,v)\equiv \nu\, n[w](x)\left[ F(v)+\hbar^2 F^{(2)}(x,v)\right] \,.
$$
Observe that such expression for $\Omega w$ can be seen as an $O(\hbar^2)-$correction
to the classical product $n(x)F(v)$.\\
In conclusion, we write Eq.~\eqref{eq:Wigeq_adim} 
in the abstract form
\begin{equation}
\left\{\!\!\!\!\!\!\!\!
\begin{array}{lcl}
&& \ze \,\displaystyle \frac{dw}{dt} = \ze \,{S} w + {\mathcal A}w +\calc w ,\\[-2mm]
\\
&& \lim_{\,t \to 0^+} \|w(t)-w_0\|_{{ X}_k} = 0
\end{array}
\right.                                          \label{system}
\end{equation}
where $w_0$ is the initial condition.\\
\indent In next Lemma we specify under which assumptions the abstract definition \eqref{eq:abstractoperators} of the operator ${\mathcal A}+\calc$ is well-posed.
\begin{lemma}   
\label{lemma:pseudo}
If $V\in H^{{k}}_x$ with $d$-admissible $k$ and $\Delta V \in L^{\infty}_x$, then the operator ${\mathcal A}+\calc$ is well-defined from $X_k$ into
itself, and is bounded by
$$
\| {\mathcal A}+\calc\|_{{\cal B}({X}_k)}\leq C(d,k)\left[\|V\|_{H^{{k}}_x}+\nu \|\Delta V\|_{L^{\infty}_x}\|F\|_{{X}_{k+2}^v}+\nu\|F\|_{{X}_{k}^v}+\nu\right].
$$
Moreover, ${\mathcal A}+\calc$ is well-defined from $X_k^v$ into itself,
and is bounded by
\begin{equation}
\label{eq:pseudo_v}
\| {\mathcal A}+\calc\|_{{\cal B}({X}_k^v)}\leq C(d,k)\left(\|V\|_{H^{k}_x}
+\nu |\Delta V(x)|\|F\|_{{X}_{k+2}^v}+ \nu\|F\|_{{X}_{k}^v}+ \nu \right)\,.\\
\end{equation}
\end{lemma}
\proof.
Here and in the following we indicate with $C$ non necessarily equal constants.\\
The arguments are similar to those in \cite{M}, so we just give a sketch of the proof.
First of all, by the product shape of the pseudo-differential operator in Fourier variables (cf.~\eqref{eq:product}), for all $w\in {X}_k$, it holds
\begin{eqnarray*}
\|\Theta[V]w\|_{X_k}^2&=& C\|\delta V{\cal F}w\|_{L^{2}_{x,\eta}}^2\! + C\left\|\sum_{i=1}^d\frac{\partial^k}{\partial{\eta_i}^k}\left(\delta V{\cal F}w\right)\right\|_{L^{2}_{x,\eta}}^2\\
&\leq& 2C\|V\|_{L^{\infty}_x}^2\|w\|_{L^{2}_{x,v}}^2\!\!\! + C\left\|\sum_{i=1}^d\frac{\partial^k}{\partial{\eta_i}^k}\left(\delta V{\cal F}w\right)\right\|_{L^{2}_{x,\eta}}^2,
\end{eqnarray*}
since $2k>d$ guarantees that $H^k_x \hookrightarrow L^{\infty}_x.$ Here, the constant $C$ is due to the Fourier transform. Then, by applying the product-formula rule and using Sobolev embeddings for the functions $V$ and ${\cal F}w\in L^2_x\otimes H^k_{\eta}$, it follows that $\|\Theta[V]\|_{{\cal B}(X_k)}\leq C \|V\|_{H^{k}_x}$.
Moreover, for all $w\in X_k$ with $2k>d$,
\begin{eqnarray}
&&\|\O w\|_{X_k}^2
\leq\nu \int_{\R^{2d}}\!\!\!(1+|v|^{2k})\left(1+\frac{\beta^4\hbar^4}{24^2m^2}|\Delta V|^2(x)\right)|F(v)|^2
\left|\int_{\R^{d}}\!\!w(x,v^{\prime})\, dv^{\prime}\right|^2\!dx\, dv \nonumber\\
&&+\int_{\R^{2d}}\!\!\!(1+|v|^{2k})\frac{\beta^6\hbar^4}{24^2m^4}
 \!\left|\sum_{r,s=1}^dv_rv_s\frac{\partial^2 V(x)}{\partial x_r\partial x_s} F(v)\right|^2\!
 \left|\int_{\R^{d}}\!\!w(x,v^{\prime})\, dv^{\prime}\right|^2\!\!dx\, dv\nonumber \\
&&\leq
\nu\,C(1+\|\Delta V\|_{L^{\infty}_x}^2) \|F\|_{X^v_k}^2 \|w\|_{X_k}^2
 + \nu \,C \|\Delta V\|_{L^{\infty}_x}^2\|F\|_{X^v_{k+2}}^2\|w\|_{X_k}^2\,,
\label{eq:stimaomega}
\end{eqnarray}
since $F\in X_k\,, \,\forall\,k$ . Then the estimate of $\| \calc\|_{{\cal B}(X_k)}$ is straightforward.
The estimate \eqref{eq:pseudo_v} in $X_k^v$
can be proved analogously. \qed \\

\noindent
We remark that the existence and uniqueness of a solution in $X_k$ of the initial value system
(\ref{system}) for any $\ze >0$ can be stated under the assumptions of Lemma \ref{lemma:pseudo} by using arguments of semigroup theory, analogously to \cite{MB}.

\section{Well-posedness of the problem with $\mathbf{\ze=0}$}
\setcounter{equation}{0}
The aim of this paper is to perform an asymptotic analysis of the system \eqref{system}, by using a Chapman-Enskog type procedure. The first step of the analysis is to solve Eq.~\eqref{system} with $\ze=0$. This corresponds to individuate the Wigner function describing the state of the system in case the interaction of the environment and the action of the potential are dominant with respect to the transport.
We remark that the function $w_{\mathrm{eq}}$ defined by \eqref{eq:wig_eq_approx} describes the state to which the system relaxes under the sole interaction with the environment.\\
We consider the equation
$({\mathcal A}+\calc) w=0 $ in the space $X_k$: the variable $x$ can be considered as a parameter in the analysis,
thus we shall study $({\mathcal A}+\calc) w=0 $ in the space $X^v_k$ for any fixed $x\in \R^d$. However, with an abuse of language, we shall indicate the operators with the same letters also when $x$ is fixed.
We can state the following proposition.
\begin{proposition}
\label{prop:kernelC}
If $V\in H^{\tilde{k}}_x$ with $\tilde{k}=\max\{2,k\},$ and $d$-admissible $k$, then
for a fixed $x\in \R^d$
\begin{equation}
\label{eq:char_ker}
{\ker}({\mathcal A}+\calc) \: = \:\{c M(v), c\in \R\}\subset X_k^v,
\end{equation}
with
\begin{equation}
\label{eq:def_M}
M(x,v):=\nu {\cal F}^{-1}\left\{\frac{{\cal F}F(\eta)}{\nu - i\delta V (x,\eta)}
 \left(1-\frac{\beta\hbar^2}{24m^2}\sum_{r,s=1}^d\eta_r\eta_s\frac{\partial^2 V(x)}{\partial x_rx_s}\right)\right\}(x,v)\,,\;v\in\R^d
\end{equation}
for any fixed $x$.
Moreover, for all $h\in X_k^v$,
$ ({\mathcal A}+\calc) w= h$ has a solution if and only if
\begin{equation}
\label{eq:cond}
\int_{\R^d}\!\!\!h(v)\, dv\;\: = \;\:0\,.
\end{equation}

\end{proposition}
\begin{remark}
{\em It can be immediately deduced by the characterization \eqref{eq:char_ker} that the solution of the equation $({\mathcal A}+\calc) w=0 $ in $X_k$ is unique, except for a factor of the sole $x$.}
\qedrem\\
\end{remark}
\proof.
By definition,
$$
{\ker}({\mathcal A}+\calc):=\{w\in X_k^v\,|\,({\nu}-\Theta[V])w=\O w\}.
$$
For all $h\in X_k^v$, the Fourier-transformed version of $({\nu}-\Theta[V])w=h$ reads
$(\nu - i\delta V){\cal F} w= {\cal F} h$. Thus,
\begin{equation}
\label{eq:inverseTheta}
w(v)=({\nu}-\Theta[V])^{-1}h(v):={\cal F}^{-1}\left(\frac{{\cal F}h(\eta)}{\nu - i\delta V (\eta)}\right)(v)
\end{equation}
is the unique solution; equivalently, the operator $(\nu-\Theta[V])$ is invertible in $X_k^v$ with
bounded inverse, defined by 
\eqref{eq:inverseTheta}.
Precisely,
\begin{eqnarray*}
\|w\|_{X_k^v}^2\!
\!\!\!\!&=&\!\!\!\!
\|({\nu}-\Theta[V])^{-1}h\|_{X_k^v}^2=C\!\!\int\!\!
\frac{|{\cal F} h(\eta)|^2} {\nu^2 +|\delta V (\eta)|^2}\, d\eta + C
\sum_{r=1}^d\int\!
\left |\frac{\partial^k}{\partial{\eta_r^k}}
\frac{{\cal F} h(\eta)}{\nu - i\delta V (\eta)}\right|^2\! d\eta\\[2mm] &\leq&\!\!\!\!\frac{C}{\nu^2}\|h\|_{L^{2}_{v}}^2+C\sum_{r=1}^d\int\!
\left |\frac{\partial^k}{\partial{\eta_r^k}}\, \frac{{\cal F}h(\eta)(\nu+ i\delta V (\eta))}{\nu^2 +(\delta V)^2 (\eta)}\right|^2d\eta\,,
\end{eqnarray*}
then, by applying product formula, it can be checked that, if $2k>d$,
\begin{equation}
\label{eq:invTheta}
\|({\nu}-\Theta[V])^{-1}\|_{{\cal B}(X_k^v)}\leq C(1+\|V\|_{H^{k}_x})\,.
\end{equation}
Then,
\begin{multline}
\label{eq:eq}
w\in \ker({\mathcal A}+\calc)\, \Leftrightarrow \, w=({\nu}-\Theta[V])^{-1}\O w\,\\[2mm]
\Leftrightarrow\, w = \nu n[w]({\nu}-\Theta[V])^{-1}(F+\hbar^2F^{(2)}[V])\,\Leftrightarrow\, w=n[w]M(v)\,,
\end{multline}
by definition of the operators $({\nu}-\Theta[V])^{-1}$ and $\O$.
From this follows the characterization \eqref{eq:char_ker} of $\ker({\mathcal A}+\calc)$, with the function $M$ defined by
$$
M(x,v):=\nu ({\nu}-\Theta[V])^{-1}(F(v)+\hbar^2F^{(2)}[V](x,v))\,\quad \forall\, v\in \R^d,
$$
with the fixed $x\in \R^d$. Since $F+\hbar^2F^{(2)}[V]\in X_k$ for all $k\in \N$, provided $\Delta V\in L^2_x$; then, due to the assumption on $V$ and to \eqref{eq:invTheta}, $M\in X_k^v$ if $2k>d$.
For all $h\in X_k$, solving $ ({\mathcal A}+\calc) w= h$ is equivalent to
$ ({\cal I}-({\nu}-\Theta[V])^{-1}\O)w=-({\nu}-\Theta[V])^{-1}h$.
Moreover, by the equivalence \eqref{eq:eq}, $\ker({\mathcal A}+\calc)= \ker({\cal I}-({\nu}-\Theta[V])^{-1}\O)$.
Since $\ker({\mathcal A}+\calc)\neq \{0\}$ , the operator ${\cal I}-({\nu}-\Theta[V])^{-1}\O$ is not injective.
If the operator $({\nu}-\Theta[V])^{-1}\O$ is compact, by the Fredholm alternative, this is equivalent to $R(({\nu}-\Theta[V])^{-1}\O)\neq X_k^v$.
The equation $({\cal I}-({\nu}-\Theta[V])^{-1}\O)w=M$ has indeed no solution, since
$$
\int_{\R^d}\!\!\!({\cal I}-({\nu}-\Theta[V])^{-1}\O)w(v)\, dv\;\: = \;\:0\,,\quad\forall\, w\in X_k^v\,,
$$
(by the definition of the operator $({\nu}-\Theta[V])^{-1}\O$), while, instead, $\int M(v) dv=\int F(v) dv=1.$
Analogously, for all
$u\in \ker({\mathcal A}+\calc), ({\cal I}-({\nu}-\Theta[V])^{-1}\O)w=u$ has no solution.\\
In conclusion, if we show that $({\nu}-\Theta[V])^{-1}\O$ is a compact operator, then we can conclude by the Fredholm alternative
that $({\cal I}-({\nu}-\Theta[V])^{-1}\O)w=h$ has a solution iff $\int h(v)\, dv = 0$.
Analogously to Lemma 1 in \cite{poupaud}, it can be constructed by Rellich-Kondrachov theorem,
a sequence of bounded finite rank operators converging to $({\nu}-\Theta[V])^{-1}\O$.
Thus the thesis follows.
\qed

\noindent Finally, let us compute the first and the second moments of the function $M$:
\begin{lemma}
\label{prop:M_prop}
Let $V\in H^{k+2}_x$ with $d$-admissible $k$. Then, the function $M$ defined by
\eqref{eq:def_M} satisfies
\begin{eqnarray}
\label{eq:eq_bis}
({\cal A}+{\cal C})w=0 &\Leftrightarrow& w=n[w]M\quad \hbox{\rm with}\quad\int\! M(x,v)\,dv = 1 \,,\\
\label{eq:first_mom}
\int\! vM(x,v)\,dv &=& -\frac{1}{\nu m}\nabla V(x)\,,\\
\label{eq:second_mom}
\int\! v\otimes v M(x,v)\,dv &=&\frac{{\mathcal I}}{\beta m}+\frac{2}{\nu^2m^2}{\nabla V\otimes \nabla V}
+\frac{\beta\hbar^2}{12m^2}{\nabla \otimes \nabla V}\,.
\end{eqnarray}
\end{lemma}
{\bf Proof}.
\eqref{eq:eq_bis} follows by \eqref{eq:eq}. Moreover, since $V\in H^{k+2}_x$ with $2k>d$, 
then the function $M$
$$
M(x,v)=\nu{\cal F}^{-1}\left(\frac{{\cal F} (F+\hbar^2F^{(2)})}{\nu-i\delta V (x,\eta)}\right)(x,v)\,,
$$
belongs to
$X_{k+2}.$
By calculus rules in the Fourier space, since $F$ is smooth, it holds
$$
\int\! vM(x,v)\,dv \;\:=\;\: i\nu \left[ \nabla_{\eta}\left(\frac{{\cal F} (F+\hbar^2F^{(2)})}{\nu-i\delta V (x,\eta)}\right)\right](x,0).
$$
By performing the derivative and then taking into account that
$$
{\cal F}(F+\hbar^2F^{(2)})(x,0)
={\cal F} F(0)=1,
\qquad
\nabla_{\eta}{\cal F}(F+\hbar^2F^{(2)})(x,0)
=0\,,
$$
and that $(\nabla_{\eta}\delta V)(x,0)=\nabla_{x}V(x)/m$, one gets
\eqref{eq:first_mom}.\\
Analogously, the second moments of $M$ are well defined, and, by calculus rules, it holds
\begin{eqnarray*}
\int\!v_iv_jM(x,v)\, dv&=&-\nu\left[\frac{\partial^2}{\partial \eta_i\partial\eta_j}\frac{{\cal F} (F+\hbar^2F^{(2)})}{\nu-i\delta V}\right](x,0)\,,\quad \forall\, i,j=1,\ldots d\,,
\end{eqnarray*}
and
\begin{eqnarray*}
-\frac{\partial^2}{\partial \eta_i\partial\eta_j}\left(\frac{{\cal F} F}{\nu-i\delta V}\right)(x,0)&=&
-\frac{1}{\nu}\left(\frac{\partial^2{\cal F} F}{\partial \eta_i\partial\eta_j}\right)(x,0)
+ \frac{2}{\nu^3}\left(\frac{\partial\delta V }{\partial\eta_i}\frac{\partial\delta V }{\partial\eta_j}\right)(x,0)\\[1.5mm]
&=&\frac{1}{\nu\beta m}+\frac{2}{\nu^3m^2}\frac{\partial V(x)}{\partial x_i}\frac{\partial V(x)}{\partial x_j}\,,\\[2mm]
-\frac{\partial^2}{\partial \eta_i\eta_j}\left(\frac{{\cal F} F^{(2)}}{\nu-i\delta V}\right)(x,0)
&=& -\frac{1}{\nu}\left(\frac{\partial^2{\cal F} F^{(2)}}{\partial \eta_i\partial\eta_j}\right)(x,0)
\;\:=\;\:\frac{\beta}{12m^2\nu}\frac{\partial^2V(x) }{\partial x_i\partial x_j}\,.
\end{eqnarray*}
Thus the thesis follows.
\qed
\begin{remark}
\label{rem:moments}
\em{
The state that describes the system under the effect of 
the interaction with the environment and of the
strong potential is described by the function $n M$, with $M$ defined by
\eqref{eq:def_M}. The fluid velocity relative to such state is non-null and given by
\eqref{eq:first_mom}. In contrast, the velocity of the system in the state $w_{\mathrm eq}$
defined by \eqref{eq:wig_eq_approx} (i.e., when it is subject to the sole influence of the environment), is $\int v\, w_{\mathrm eq}\, dv=0$,
as expected since it is an equilibrium state.
Moreover, the expression of the second moment tensor \eqref{eq:second_mom} has to be compared with
$$
\int v\otimes v \,w_{\mathrm eq} \,dv=n\left(\frac{ {\mathcal I}}{\beta m}+\frac{\beta\hbar^2 }{12m^2}
{\nabla \otimes \nabla V}\right).
$$
They differ by the second summand in \eqref{eq:second_mom} that
is to be referred to the strong-field assumption (cf.~\cite{poupaud}).
}\qedrem\\
\end{remark}

\noindent As a consequence of Proposition \ref{prop:kernelC}, the following subspace is well-defined
$$
\left(X_k\right)_M \::=\:\{ \alpha(x)M(x,v), \alpha\in  L^2_x\}\subset X_k\,,
$$
which coincides with $\ker (\cal{A}+\cal{C})$ when $\cal{A}+\cal{C}$ is considered as an operator on $X_k$. Accordingly, we can
decompose the space $X_k$ as
\begin{equation}
\label{eq:decomposition}
X_k = \left(X_k\right)_M \oplus \left(X_k\right)^0
\end{equation}
with
$$
\left(X_k\right)^0:=\left\{ w\in X_k \left| \int\!\! w(x,v)\,dv = 0 \right.\right\}\,,
$$
and define the corresponding spectral projection $\cal P$ from
$X_k$ into $\left(X_k\right)_M$, by
\begin{equation}
{\cal P}w := M \int_{\R^d_v}\!\! w(x,v)\,dv\,,   \nonumber
\end{equation}
and ${\cal Q} := {\cal I} - {\cal P}$. The following corollary is still a preliminary result for our asymptotic procedure.
\begin{corollary}
\label{lemma1} 
Let $V\in H^{k}_x$ with $d$-admissible $k$. Then, the operator ${\cal Q(A+C)Q} $ is an isomorphism of $(X_k)^0$ onto itself, with
\begin{equation}
\label{eq:normQACQ}
\| {\mathcal A}+\calc\|_{{\cal B}({X}_k)}\leq C(d,k)\left(\|V\|_{H^{{k}}_x}+\nu\right).
\end{equation}
If, in addition, $V\in H^{k+j}_x$ with $j>0$, then ${\cal Q(A+C)Q} $ is an isomorphism of $(H^j_k)^0$ onto itself, with
\begin{equation}
\label{eq:normQACQH}
\| {\mathcal A}+\calc\|_{{\cal B}(H^j_k)}\leq C(d,k,j)\left(\|V\|_{H^{{k+j}}_x}+\nu\right).
\end{equation}
\end{corollary}
{\bf Proof.}
The operator ${\cal Q(A+C)Q}$, when considered    
as an operator acting on $(X_k)^0$, reduces to
\begin{equation}
\label{eq:qacq}
{\cal Q(A+C)Q} u = \Theta[V]u - \nu u, \quad \forall \, u \in (X_k)^0.
\end{equation}
Then the thesis follows from Lemma~\ref{lemma:pseudo},
Prop.~\ref{prop:kernelC}, and by the skew-simmetry of the pseudo-differential operator.
The second statement and estimate \eqref{eq:normQACQH} can be proved analogously.
\qed\\
\section{The asymptotic expansion}
\label{sec_cae}
\setcounter{equation}{0}
According to the decomposition of the space $X_k$, every function $w \in X_k$ can be written as
$w={\cal P}w+{\cal Q}w$, with ${\cal P}w\in(X_k)_M$ and ${\cal Q}w\in(X_k)^0$. Let us call $ \zf:= {\cal P}w $ and $\psi:={\cal Q}w $.
Observe that, for all $w\in X_k, \int {\cal P}w(x,v)\, dv =  n[w](x)$, while $\int {\cal Q}w(x,v)\, dv = 0$, that is, we separate the part of $w$ that contributes to the density $n[w]$ from the other one. Precisely, it holds ${\cal P}w= n[w]M$, by definition.\\
\indent Applying formally the projection $\cal P$, respectively
$\cal Q,$ to the Wigner-BGK equation (\ref{system}) with unknown $w$,
we obtain the following system of equations with unknown $\zf$ and $\psi$
\begin{equation}
\left\{
\begin{array}{rcl}
\displaystyle{\frac{\p \zf}{\p t}} &=&
\psp \zf + \psq \psi \\[3mm]
\displaystyle{\frac{\p \psi}{\p t}} &=&
\qsp \zf + \qsq \psi + \displaystyle \frac{1}{\ze} {\cal Q}({\cal A}+{\cal C}){\cal Q} \psi
\end{array}
\right.                                                   \label{sispro}
\end{equation}
where we used  $({\cal A}+{\cal C})\calp \zf =0$ and ${\calp ({\cal A}+{\cal C}) \calq} \psi=0$, together with the initial conditions
\begin{equation}
\zf(0) = \zf_0 = \calp w_0\,, \qquad
\psi(0)= \psi_0 = \calq w_0.
                                                \label{icpro}
\end{equation}
System \eqref{sispro} 
consists of an evolution problem with unknown functions $\zf= n[w]M$ and $\psi$, and it is supplemented by the initial conditions \eqref{icpro}. It is a reformulation of (\ref{system}).
\\
Since we expect the solution $w$ to be subject to rapid changes for small times, we split the functions $\zf$ and $\psi$ into the sums of the ``bulk'' parts $\baf$
and $\bar \psi$ and of the ``initial layer'' parts $\tilde \zf$ and $\tilde \psi$,
\begin{equation}
\zf(t) =\fibar(t) + \tilde \zf \left(\frac{t}{\ze}\right)\,,\qquad
\psi(t)=\bar \psi(t)+ \tilde \psi\left(\frac{t}{\ze}\right).   \nonumber
\end{equation}
The bulk part $\bar\zf$ is left {\em unexpanded} and the other parts
are expanded in terms of $\ze$ as follows
\begin{eqnarray}
\label{expansion}
\fitil(\zt) &=& \fitil_0(\zt) + \ze \fitil_1(\zt) + \ze^2 \fitil_2(\zt)
+\ldots \nonumber\\[2mm]
\bar \psi (t)   &=& \psibar_0(t) + \ze \psibar_1(t)+ \ze^2 \psibar_2(t)
+ \ldots \\[2mm]
\psitil(\zt)&=& \psitil_0(\zt) + \ze \psitil_1(\zt)+ \ze^2 \psitil_2(\zt)
+ \ldots ,\nonumber
\end{eqnarray}
with $\zt={t}/{\ze}.$
Accordingly, Eqs.~\eqref{sispro} for the bulk part terms of the expansion up to the order $\ze^2$ become
\begin{equation}
\left\{
\begin{array}{rcl}
\displaystyle{
\frac{\p \bar \zf}{\p t}} &=&
\psp \bar \zf + \psq \bar\psi_0 + \ze \psq \bar\psi_1 \\[2mm]
0&=&{\cal Q}({\cal A}+{\cal C}){\cal Q}\bar \psi_0\\[2mm]
0&=&\qsp \bar\zf +{\cal Q}({\cal A}+{\cal C}){\cal Q} \bar\psi_1\,
\end{array}
\right.                                                   \label{sisprobis}
\end{equation}
while the equations for the initial layer parts read
\begin{equation}
\left\{
\begin{array}{rcl}
\displaystyle{\frac{\p \tilde \zf_0}{\p \zt}}&=& 0,\\[2mm]
\displaystyle{\frac{\p \tilde \zf_1}{\p \zt}}&=&
                {\cal P}S{\cal Q} \tilde \psi_0 (\zt)\\[2mm]
\displaystyle{\frac{\p \tilde \psi_0}{\p \zt}}&=&
                 {\cal Q(A+C) Q} \tilde \psi_0(\zt)\\[2mm]
\displaystyle{\frac{\p \tilde \psi_1}{\p \zt}}&=&
{\cal Q(A+C) Q} \tilde \psi_1(\zt)+{\cal Q}S{\cal Q} \tilde \psi_0 (\zt)
\end{array}
\right.                                                        \label{initial}
\end{equation}
and the initial conditions \eqref{icpro} yield
\begin{equation}
\left\{
\begin{array}{rcl}
\fibar(0)+  \tilde \zf_0(0)+ \ze \tilde \zf_1(0)&=&\zf_0 \\
     \bar \psi_0(0)+  \tilde \psi_0(0)&=&\psi_0  \\
   \bar \psi_1(0) +  \tilde \psi_1(0) &=&0 \, .
\end{array}
\right.                                                        \label{cisy}
\end{equation}
System \eqref{sisprobis}, together with \eqref{initial}-\eqref{cisy}, is an $\mathcal{O}(\ze^2)$-approximated version of \eqref{sispro} with \eqref{icpro}, once the expansion \eqref{expansion} has been introduced. 
In fact, the equations in \eqref{sisprobis} can be decoupled: by Corollary \ref{lemma1}, the operator ${\cal Q}({\cal A}+{\cal C}){\cal Q}$ is invertible in $(X_k)^0$, thus
\begin{eqnarray}
\label{eq:psibar0}
\psibar_0 &\equiv& 0
\\
\label{eq:psibar1}
\psibar_1 &=& -({\cal Q}({\cal A}+{\cal C}) \calq)^{-1} \qsp \fibar\,,   
\end{eqnarray}
which implies
\begin{equation}
\label{diffu1}
\displaystyle{\frac{\p \fibar}{\p t}} = \psp \fibar
- \ze \psq ({\cal Q}({\cal A}+{\cal C}) \calq)^{-1} \qsp \fibar\,.
\end{equation}
Thus, system \eqref{sisprobis} reduces to the system \eqref{eq:psibar1}-\eqref{diffu1}, with unknown functions $\fibar(x,v,t)= n(x,t) \ M(x,v)$ and $\psibar_1$. Next section shall be dedicated to reformulate Eq.~\eqref{diffu1} as an equation with unknown $n$. The analysis of system \eqref{initial}, with unknown $\fitil$ and $\psitil$ and initial conditions \eqref{cisy}, is postponed to Section \ref{sec:initiallayer}: it shall provide an appropriate initial condition for Eq.~\eqref{diffu1}. Finally, in Sections 7 and 8 we shall establish a well-posedness result for the approximated problem.
In our main theorem (cf.~Thm.~\ref{maintheorem}), we shall prove that the solution $\varphi+\psi$ of equations \eqref{sispro} indeed differs from $[\fibar(t)+ \fitil_0(\zt) + \ze \fitil_1(\zt)]+[\psibar_0(t) + \ze \psibar_1(t)+ \psitil_0(\zt) + \ze \psitil_1(\zt) ]$, satisfying the approximated problem \eqref{sisprobis}-\eqref{cisy}, by a term of order $\ze^2$.

\section{The high-field quantum drift-diffusion equation}
\label{strongfield}
\setcounter{equation}{0}
The aim of the present section is the reformulation of the abstract equation \eqref{diffu1} as an equation with unknown $n$.
\begin{lemma}
\label{lemma:ourQDD}
Let $V\in H^{k+2}_x$ with $d$-admissible $k$. Eq.~\eqref{diffu1} with unknown $\baf(x,v,t) = n(x,t) M(x,v)$
can be rewritten as an evolution equation with unknown $n(x,t)$ of the form
\begin{eqnarray}
\label{eq:ourQDD}
\frac{\partial n}{\partial t} &-&\frac{1}{\nu m}\nabla  \cdotp (n\nabla V)
-\frac{\ze}{\nu\beta m} \nabla \cdotp \nabla  n\nonumber\\
&-&\frac{\ze}{\nu^3m^2} \left [\nabla  \cdotp (n (\nabla \otimes \nabla) V\nabla V) + \nabla \cdotp\nabla \cdotp (n{\nabla V \otimes \nabla V})
 \right]\nonumber\\
&-&\frac{\ze\beta\hbar^2}{12\nu m^2}\nabla  \cdotp \nabla \cdotp\left(n \nabla \otimes \nabla V \right)=0
\end{eqnarray}
\end{lemma}
\begin{remark}
\em{
The first line of \eqref{eq:ourQDD} consists of the terms of the classical DD equation.
The second line is peculiar of the strong-field assumption, being a correction of order $\ze$, and consists of the additional term
$$
\frac{1}{\nu}\frac{\nabla V\otimes \nabla V}{\nu^2m^2}
$$
in the pressure tensor, and of the term
$$
\frac{1}{\nu}
\left(\frac{\nabla\otimes\nabla) V \nabla V} {\nu^2m^2}\right)
$$
in the drift term. Both terms are quadratic in the potential $V$.
The second line can also be written as
$$
-\frac{\ze}{\nu^3m^2} \nabla  \cdotp [\nabla V \otimes \nabla V \nabla n + n \left(2\nabla\otimes\nabla V\nabla V
+ \Delta V\nabla V\right)]\,.
$$
This expression is the same obtained in \cite{poupaud} from the semi-classical Boltzmann equation with high-field scaling.
The last line is the quantum pressure term (cf.~\cite{Gard94,gasmar}).
}\qedrem\\
\end{remark}
The proof requires the following preliminary lemmata.
\begin{lemma}
\label{lemma:D2}
Let $V\in H^{k+2}_x$ with $d$-admissible $k$. 
Then the equation
\begin{equation}\label{eqD2}
({\cal A}+\calc) w=   M\left(-v + \int\!\! v M\, dv \right)\,,
\end{equation}
admits a unique solution $(D_2)_i \in (X_{k+1})^0\,, \forall\, i=1\,,\ldots d$.
Moreover, let ${\sf D}$ be the matrix defined by
$$
{\sf D}_{ij} (x):= \int\! v_i (D_2)_j(x,v)dv,
$$
then
\begin{equation}
{\sf D}(x)=\frac{1}{\nu}\left(\frac{{\mathcal I}}{\beta m}+\frac{1}{\nu^2m^2}{\nabla V\otimes \nabla V}
+\frac{\beta\hbar^2}{12m^2}\nabla\otimes\nabla V\right)(x)\,.
\label{diftensor}
\end{equation}
\end{lemma}
{\bf Proof}.
Since the right-hand side of Eq.~\eqref{eqD2} belongs to $(X_{k+1})^0$, it satisfies the compatibility condition \eqref{eq:cond} and there exist $(D_2)_i \in (X_{k+1})^0\,, \forall i=1\,,\ldots d$ satisfying \eqref{eqD2}. More explicitly, $D_2$ solves
\begin{equation}
\label{eq:eqD2}
(\Theta[V]-\nu) D_2(x,v) = -M(x,v)\left[ v+  \frac{\nabla V(x)}{\nu m} \right],
\end{equation}
since $\int D_2(v)dv = 0$ and by \eqref{eq:first_mom}. Multiplying the left-hand side of Eq.~\eqref{eq:eqD2} by $v+\nabla V/(\nu m)$ and integrating over
${\R^d}$ we obtain
$$
\int\! \left(v+\frac{\nabla V(x)}{\nu m}\right)\otimes\left[ \Theta[V]-\nu \right ] D_2(x,v)\, dv
= - \nu\int\!\! v_i (D_2)_j(x,v)\,dv\,,
$$
by using the skew-simmetry and $(D_2)_i\in (X_{k+1})^0.$
Thus Eq.~\eqref{eq:eqD2} gives
\begin{eqnarray}
\label{eq:matrix}
\nu \, {\sf D}_{ij} (x) &=&
\int\! \Big(v+\frac{\nabla V(x)}{\nu m}\Big)\otimes \Big(v+\frac{\nabla V(x)}{\nu m}\Big)M(x,v)\,dv\\
&=& \int\! v\otimes vM(x,v)\,dv- \frac{1}{\nu^2 m^2}\nabla V(x)\otimes\nabla V(x)\,.\nonumber
\end{eqnarray}
From \eqref{eq:second_mom}, the thesis follows.
\qed
\begin{remark}
{\em
By considering the expression \eqref{eq:first_mom} for the fluid velocity, we can recognize in \eqref{eq:matrix} the classical
definition of the pressure tensor in terms of $M$. This is to be expected, since the function $M$ is the solution of the the evolution problem with $\ze=0$. Thus, the term with diffusion tensor ${\sf D}$ is what we expected to find as correction of first order in $\ze$.
By \eqref{diftensor}, it consists of the standard temperature and quantum pressure tensors, and of the additional
tensor ${1}/({\nu^3 m^2})\nabla V(x)\otimes\nabla V(x)$, to be referred to the strong-field assumption (cf.~Remark \ref{rem:moments}).
}\qedrem\\
\end{remark}
\begin{lemma}
\label{lemma:D1}
Let $V\in H^{k+2}_x$ with $d$-admissible $k$. The following equation
\begin{equation}\label{eqD1}
({\cal A}+\calc)w=-v\cdotp \nabla_x  M+M\int\!\! v \cdotp \nabla_xM\, dv\,,
\end{equation}
admits a unique solution $D_1 \in (X_{k+1})^0\,.$ Moreover, the vector $W$ defined by
$$
{\sf W}(x):=\int v D_1(x,v)\,dv
$$
can be calculated explicitly
\begin{equation}
\label{eq:W}
{\sf W}(x) = \frac{1}{\nu}\left(2\frac{\nabla\otimes \nabla V}{\nu^2 m^2}\nabla V(x) +\frac{\Delta V \nabla V}{{\nu^2 m^2}}
+\frac{\beta\hbar^2}{12m^2} \nabla \cdotp \nabla\otimes \nabla V\right)(x) \,.
\end{equation}
\end{lemma}
{\bf Proof}.
Under the regularity assumptions on $V$, $M\in H_{k+1}^1:= H^1_x\otimes X_{k+1}^v$ and the right-hand side of Eq.~\eqref{eqD1} belongs to $(X_{k+1})^0$, thus it exists $D_1\in (X_{k+1})^0$ solving
$$
(\Theta[V]-\nu) D_1(x,v) = -v\cdotp\nabla_x M(x,v)- M(x,v)\nabla \cdotp\frac{\nabla V(x)}{\nu m}\,,
$$
which is equivalent to \eqref{eqD1}, since $\int D_1(v)\,dv = 0$.
Multiplying
by $v+{\nabla V}/{\nu m}$ and integrating over
${\R^d}$ we obtain
\begin{eqnarray}
\label{eq:vD_2}
\nu\int\!\! v D_1(x,v)\,dv &=& \int\!\! \left(v +\frac{\nabla V(x)}{\nu m}\right)\nabla_x \cdotp\left(v +\frac{\nabla V(x)}{\nu m}\right)M(x,v)\, dv \nonumber\\
\label{eq:vD_2bis}
&=& \nabla_x\cdotp\int\!\!v \otimes v M(x,v)\, dv - \frac{1}{\nu^2m^2}\Delta V(x)\nabla V(x)\,.
\end{eqnarray}
The thesis follows directly from \eqref{eq:second_mom} and
$$
\nabla \cdotp (\nabla V\otimes\nabla V)= \Delta V\nabla V+\left(\nabla\otimes\nabla\right)V\nabla V\,.
$$
\qed

\noindent{\bf Proof of Lemma 6.1} First of all, let us write the explicit expression of the operators appearing in Eq.~\eqref{diffu1}. Observe that $V\in H_x^{k+2}$ implies $M\in H_{k+1}^1$. By definition,
\begin{equation}
\label{eq:PSP}
\psp \fibar = -M \int\!\! v\cdotp \nabla_x (n M)\, dv
=-\ M\nabla_x \cdotp\left( n \int\!\! v M\, dv \right)\,, \nonumber
\end{equation}
and $({\cal P}S{\cal Q } g)=({\cal P}Sg)- ({\cal P}S{\cal P} g)$, i.e., explicitly,
\begin{equation}
\label{eq:psq}
({\cal P}S{\cal Q } g)=- M
\left(\int\!\! v\cdotp \nabla_x g\,dv - \int\!\!\nabla_x g\,dv\,\cdotp \int\!\! v\, M\,dv -  \int\!\!g\,dv\nabla_x \cdotp \int\!\! v M dv \right)\,.
\end{equation}
Moreover $\qsp \fibar =( S {\cal P}- \psp)\fibar$, i.e., explicitly,
\begin{equation}
\label{eq:QSP}
\qsp \fibar = n\left[-v\cdotp \nabla_x  M+M\int\!\! v \cdotp \nabla_xM dv \right]
+\nabla_x n\cdotp
\left[ M\left(-v+ \int\!\! v M dv \right) \right] .
\end{equation}
By Lemmata \ref{lemma:D2} and \ref{lemma:D1}, $D_2(x,v)\equiv(D_2)_i(x,v)$ and $D_1(x,v)$ are solutions with $(D_2)_i, D_1\in\left(X_{k+1}\right)^0$ of Eqs.~\eqref{eqD2} and \eqref{eqD1}, respectively. Then, by some manipulations,
\begin{equation}
\label{eq:standard}
\psq (\calq ({\cal A}+\calc)\calq)^{-1} \qsp \fibar = {\cal P}S (D_2 \cdotp\nabla n) + {\cal P}S (D_1 n)\,,
\end{equation}
where the right-hand side can be written explicitly as
\begin{eqnarray*}
{\cal P}S (D_2 \cdotp\nabla n) \;=& \! -M\int v\cdotp \nabla_x( D_2 \cdotp \nabla n)\,dv \!&=\;-M\nabla_x\cdotp \left[\left( \int\!\!v \otimes D_2 \,dv \right)\cdotp\nabla n\right]\,,\\
{\cal P}S (D_1 n) \:=&\! -M\int v\cdotp \nabla_x( D_1 n)\,dv\!&=\;-M \nabla_x\cdotp \left[n\int\!\!v D_1 \,dv\right]\,.
\end{eqnarray*}
Hence, by simplifying the common factor $M$, Eq.~(\ref{diffu1}) reads
\begin{eqnarray*}
\frac{\p n}{\p t} =&-&  \nabla_x \cdotp \left( n\int\!\! v M dv \right)
+\epsilon \nabla_x \cdotp \left({\sf D}\cdotp\nabla n
+n{\sf W}\right)\,,
\end{eqnarray*}
and the thesis follows by using \eqref{diftensor} and \eqref{eq:W}.
\qed\\
\noindent As a consequence of \eqref{eq:standard}, Eq.~\eqref{eq:psibar1} defining the other non-zero term of the bulk part expansion, $\bar \psi_1$, can be rewritten as
\begin{equation}
\label{eq:J_1}
\bar \psi_1(x,t)= - \left\{D_2 \cdotp \nabla n + D_1 n\right\}\,.
\end{equation}
The explicit version of this expression shall be given in Eq.~\eqref{eq:psi1explicit}.

\section{Rigorous results: the initial layer part}
\label{sec:initiallayer}
\setcounter{equation}{0}
The aim of the present section is to prove existence and regularity of the solutions of Eqs.~\eqref{initial}, together with the initial conditions (\ref{cisy}). The first equation in \eqref{initial} yields
$$
\fitil_0(\zt) \equiv 0,
$$
since we expect that $\lim_{\tau\to \infty} \fitil_0(\zt)=0.$ The equation for $\tilde{\psi}_0$ with the appropriate initial condition coming from \eqref{cisy}-\eqref{eq:psibar0}, is
\begin{equation}
\left\{
\begin{array}{ccl}
\displaystyle \frac{\p \psitil_0}{\p \tau} &=& {\cal Q(A+C)Q} \tilde{\psi}_0\\
\, \rule{0mm}{5mm}\tilde{\psi}_0 (0) &=& \psi_0\,.
\end{array}
\right.
\label{psitilde0}
\end{equation}
We recall that the operator ${\cal Q(A+C)Q}$ on $(X_k)^0$ reduces to
$$
{\cal Q(A+C)Q} w = \Theta[V]w - \nu w, \quad \forall \, w \in (X_k)^0\,,
$$
(cf.~\eqref{eq:qacq}). By the product shape in Fourier-variables of the pseudo-differential operator (cf.~\eqref{eq:product}), it is more convenient to consider the equation for ${\cal F}\tilde{\psi}_0$, that looks like
$$
\frac{\partial}{\partial{\tau}}{\cal F}\tilde{\psi}_0 (x,\eta, \tau)= (i\,\delta V(x,\eta)-\nu) {\cal F} \tilde{\psi}_0(x,\eta,\tau)\,.
$$
Thus, we define, for all $w\in L^2(\R^6;\R),$ the semigroup $G(\tau)$
\begin{eqnarray}
\label{eq:FourierG}
G(\tau)w(x,v)&:=&{\cal F}^{-1}\left(e^{(i\,\delta V(x,\eta)-\nu)\tau}
{\cal F} w(x,\eta)\right)\\
&=&e^{-\nu\tau}{\cal F}^{-1}\left(e^{i\,\delta V(x,\eta)\tau}{\cal F} w(x,\eta)\right)\,,\quad \forall\, \tau\geq 0\,.\nonumber
\end{eqnarray}
The function $\psitil_0(\tau)\equiv G(\tau)\psi_0$ formally satisfies system \eqref{psitilde0}. Moreover,
\begin{lemma}   \label{lemma2}   
If $w \in (X_k)^0$ and $V\in H_x^{k}$ with $d$-admissible $k$, then there exist ${0<\nu_k<1},$
and a constant $C(\|V\|_{H_x^k})>0$, such that
\begin{equation}
\label{eq:decay}
\|G(\tau)w\|_{X_k} \leq C (\|V\|_{H_x^{k}}) \,e^{-\nu_k\tau}\|w\|_{X_k}\,.
\end{equation}
If, in addition, $w \in (H_k^j)^0$ and $V\in H_x^{k+j}$, then
\begin{equation}
\label{eq:decayH}
\|G(\tau)w\|_{H_k^j} \leq C (\|V\|_{H_x^{k+j}}) \,e^{-\nu_{k+j}\tau}\|w\|_{H_k^{j}}\,,
\end{equation}
with appropriate $C (\|V\|_{H_x^{k+j}})>0,$ and $0<\nu_{k+j}<1.$ Eq.~\eqref{eq:FourierG} defines a strongly continuous semigroup on $(X_k)^0$ (respectively on $(H_k^j)^0$).
\end{lemma}
\proof.
By definition we have
\begin{eqnarray*}
\|G(\tau)w\|_{X_k}&\leq& C e^{-\nu\tau}\!\left(\|e^{i\,\delta V(x,\eta)\tau}
{\cal F} w(x,\eta)\|_{L^2_{x,\eta}}+\|\nabla^k_{\eta}(e^{i\,\delta V(x,\eta)\tau}
{\cal F} w(x,\eta)\|_{L^2_{x,\eta}}\right)
\\
&\leq& C e^{-\nu\tau}\!\left(\|w\|_{L^2_{x,v}}+P_k(\tau \|V\|_{H_x^{k}})\|w\|_{X_k}\right)\\
&\leq& e^{-\nu_k\tau}\max_{\tau\geq 0}\{e^{-(\nu-\nu_k)\tau}P_k(\tau \|V\|_{H_x^{k}})\}\|w\|_{X_k}\,,
\end{eqnarray*}
where $0<\nu_k<\nu$ and $P_k$ is a polynomial of degree $k$. The estimate \eqref{eq:decayH} can be proved analogously.
The last assertion follows immediately by applying Hille-Yosida Thm., thanks to \eqref{eq:decay} (respectively \eqref{eq:decayH}).
\qed\\

\noindent
With Lemmata \ref{lemma1} and \ref{lemma2}  we can prove the following proposition.
\begin{proposition}
\label{prop:estimates_initial}
If $w_0\in H_{k+1}^1$ and $V\in H_x^{k+2}$, with $d$-admissible $k$, then all terms of the initial layer expansion are well-defined and
satisfy the
following estimates:
\begin{eqnarray}
\label{eq:tilde_psi_0}
 \|\tilde\psi_0(\tau)\|_{X_k} &\leq &{\rm
M}_1e^{-\nu_{k}\tau}\|w_0\|_{X_{k}}\,,\\
\label{eq:tilde_zf_1}
 \|\tilde\zf _1(\tau)\|_{X_k} &\leq &{\rm
M}_2e^{-\nu_{k+2}\tau}\|w_0\|_{H^1_{k+1}}\,,\\
\label{eq:tilde_psi_1}
\|\tilde\psi_1(\tau)\|_{X_k} &\leq& {\rm M}_3e^{-\nu_{k+2}\tau}
\|w_0\|_{H^1_{k+1}}\,,
\end{eqnarray}
for some constants ${\rm M}_1$, ${\rm M}_2$ and ${\rm M}_3$(depending on norms of $V$).
\end{proposition}
{\bf Proof}.
The unique solution of system \eqref{psitilde0} is
\begin{equation}
\tilde\psi_0(\zt) = G(\zt)\psi_0\,,
\end{equation}
and \eqref{eq:tilde_psi_0} follows immediately from (\ref{eq:decay}) since $\psi_0={\cal P}w_0\in X_k$. Now we shall consider, among Eqs.~\eqref{initial}, the following one:
$$
\frac{\p \tilde \zf_1}{\p \zt}(\zt)=
                \psq \tilde \psi_0 (\zt)\,.
$$
The right hand side is well-defined by considering the definition of the operator ${\cal P}S{\cal Q}$ (cf.~\eqref{eq:psq}), together with Lemma \ref{lemma2}, since $\psi_0 \in (H^1_{k+1})^0$ and $V\in H^{k+2}_x$. By integrating with respect to $\tau$ and considering $\lim_{\tau\to \infty} \tilde \zf_1(\zt)=0$, we obtain
\begin{eqnarray*}
\tilde\zf_1(\tau) &=& -\int_{\tau}^{\infty}\psq \tilde\psi_0(s)\,ds \\
&=&-\int_{\tau}^{\infty}\psq [{\cal Q (A+C) Q}]^{-1}[{\cal Q(A+C)Q} ] G(s)\psi_0\,ds =\\
&=&-\psq[{\cal Q(A+C)Q}]^{-1}\int_{\tau}^{\infty}\!\!{\cal Q(A+C)Q} G(s)\psi_0\,ds.
\end{eqnarray*}
Last integral is well-defined since the integrand ${\cal Q(A+C)Q} G(s)\psi_0 $
is equal to $G(s){\cal Q(A+C)Q}\psi_0$, which is continuous in $H^1_{k+1}$ (by Lemma \ref{lemma2}).
Moreover
$\psq$ $[{\cal Q(A+C)Q}]^{-1} \in {\cal L}(H^1_{k+1},X_k)$ (by Corollary \ref{lemma1}),
then it can be taken outside the integral. Since
$$
 {\cal Q(A+C)Q} G(s)\psi_0 = \frac{\p G(s)\psi_0}{\p s}\,,
 $$
thanks to the exponential decay of $G$ in $H^1_{k+1}$ and the
continuity of the operator $\psq[{\cal Q(A+C)Q}]^{-1}$, we obtain
\begin{equation}
\label{phipre}
 \tilde\zf_1(\tau) = \psq[{\cal Q(A+C)Q}]^{-1}G(\tau)\psi_0\,,
\end{equation}
and, in particular,
$$
 \tilde\zf_1(0) = \psq[{\cal Q(A+C)Q}]^{-1}\psi_0,
  $$
 which provides the initial datum.
 Then, (\ref{eq:tilde_zf_1}) follows from the estimate
\begin{eqnarray}
\label{eq:tilde_zf_1proof}
 \|\tilde\zf _1(\tau)\|_{X_k} &\leq & |\|\psq[{\cal Q(A+C)Q}]^{-1}\|| \; \|G(\tau)\psi_0\|_{H^1_{k+1}}
\leq\\
&\leq&{\rm M}_2e^{-\nu_{k+2}\tau}\|\psi_0\|_{H^1_{k+1}}\,,\nonumber
\end{eqnarray}
where $|\|\cdot\| |$ denotes the norm in ${\cal L}(H_{k+1}^1,X_k)$.
Finally, we prove that the equation
$$
\frac{\p \tilde \psi_1}{\p \zt}(\zt)=
{\cal Q(A+C)Q}\tilde \psi_1(\zt)+{\cal Q}S{\cal Q} \tilde \psi_0 (\zt)
$$
is classically
solvable. The initial condition for $\tilde{\psi}_1$ can be obtained from Eqs.~ \eqref{cisy}, together with Eq.~\eqref{eq:psibar1} for $\bar \psi_1$,
$$
\tilde \psi_1(0)= - \bar \psi_1(0) = [{\cal Q}({\cal A}+{\cal C}) \calq]^{-1} \qsp \fibar(0)\,,
$$
and by considering
\begin{equation}
\fibar(0) = \zf_0 -  \fitil_0(0) - \ze \fitil_1(0) =
 \zf_0 - \ze \psq [{\cal Q(A+C)Q}]^{-1} \psi_0.
\label{difiv}
\end{equation}
Since $\tilde \psi_1(0)$ is by itself a correction of order $\ze$, we neglect the term of order $\ze$ in the expression for $\fibar(0)$, and it yields
$$
\tilde \psi_1(0)= [{\cal Q}({\cal A}+{\cal C}) \calq]^{-1} \qsp\zf_0\,.
$$
By Lemma \ref{lemma2}, $G$ is a semigroup on $(H^1_{k+1})^0$ and,
thanks to the assumption on $w_0$, $\psi_0$ is in the domain of ${\cal Q(A+C)Q}$
when it is defined in $D(S)\cap(H^1_{k+1})^0$. Therefore
$\tilde\psi_0(\tau) = G(\tau)\psi_0$
is differentiable on $[0,\infty[$ in $X_{k+1}$ so that the inhomogeneous term
$\qsq\tilde\psi_0(\tau)$ is differentiable on $[0, \infty[$ in $X_k$.
This, together with $\tilde\psi_1(0) = ({\cal Q(A+C)Q})^{-1}\qsp\zf_0
\in D({\cal Q(A+C)Q})$, shows that
\begin{equation}
\label{eq:tildepsidef}
\tilde\psi_1(\tau) = G(\tau)\tilde \psi_1(0) + \int_{0}^{\tau}G(\tau -
\sigma)\qsq G(\sigma)\psi_0d\sigma
\end{equation}
is a classical solution.
The estimate \eqref{eq:tilde_psi_1} follows from $[{\cal Q(A+C)Q}]^{-1} \qsp\in {\cal L}(H^1_k,X_k)$ and from
\eqref{eq:tilde_psi_0}:
\begin{eqnarray*}
\|\tilde\psi_1(\tau)\|_{X_{k}} &\leq&
{\rm K}_1e^{-\nu_k\tau}\|\zf_0\|_{X_{k}}
+ {\rm K}_2 e^{-\nu_k\tau}\int_{0}^{\tau}e^{(\nu_k-\nu_{k+2})\sigma}\|\psi_0\|_{H^1_{k+1}} d\sigma\\
&\leq& {\rm K}_1 e^{-\nu_k\tau}\|\zf_0\|_{X_{k}} + {\rm K}_3e^{-\nu_{k+2}\tau}
 \|\psi_0\|_{H^1_{k+1}} \:\;\leq \:\; {\rm M}_3e^{-\nu_{k+2}\tau}\|w_0\|_{H^1_{k+1}}\,.
 \end{eqnarray*} \qed\\
In order to obtain an initial value for Eq.~\eqref{eq:ourQDD} with unknown $\bar\varphi=n\,M$, we consider again \eqref{difiv}. Let us call $n_0(x) = \int w_0(x,v)\, dv $, such that $\zf_0=n_0M$ and, by dividing both sides of the expression \eqref{difiv} by $M$, it yields
\begin{equation}
\label{cx0}
n(x,0) = n_0(x) + \ze \int\!\! v \cdotp \nabla_x {\cal F}^{-1}\left(\frac{{\cal F}\psi_0}{i \delta V - \nu}\right)(x,v)\, dv\,,
\end{equation}
by using the explicit expression the operator $({\cal Q}({\cal A}+{\cal C}){\cal Q})^{-1}$ (cf.~\eqref{eq:qacq}).
In the following we shall call
\begin{equation}
\label{cx0b}
n(x,0) = n_0(x) + \ze n_1(x)\quad \hbox{with} \quad n_1(x):=\int\!\! v \cdotp \nabla_x {\cal F}^{-1}\left(\frac{{\cal F}\psi_0}{i \delta V - \nu}\right)(x,v)\, dv\,.
\end{equation}
The explicit expression for \eqref{phipre} can be obtained analogously and reads
\begin{equation}
\tilde{\zf}_1(\tau) = - M\int\!\! v \cdotp \nabla_x {\cal F}^{-1}\left(\frac{{\cal F}G(\tau)\psi_0}{i \delta V - \nu}\right)\, dv\,.     \nonumber
\end{equation}
\section{Well-posedness of the high-field QDD equation}
\setcounter{equation}{0}
In this section, we establish a well-posedness and regularity result for Eq.~\eqref{eq:ourQDD},
with a given external potential $V$.
The equation can be rewritten in divergence form as
\begin{equation}
\label{eq:qdd1}
 \frac{\pt n}{\pt t} - {\cal D} n - {\cal G} n - {\cal E} n =0\,,
\end{equation}
where we indicate
$$
{\cal D} n = {\ze}\nabla \cdot ({\sf D} \nabla n)\,,\quad
{\cal G} n = {\ze}\nabla \cdot ({\sf W}\,n)\,,\quad  {\cal E} n = \nabla \cdot ( {\sf E}\,n)
$$
with
\begin{eqnarray*}
{\sf D}\equiv{\sf D}(x)&:=& \frac{1}{\nu}\left(\frac{{\mathcal I}}{\beta m}+\frac{1}{\nu^2m^2}{\nabla V\otimes \nabla
V}+\frac{\beta\hbar^2}{12m^2}\nabla\otimes\nabla V\right)(x)\,,\\[2mm]
{\sf W}\equiv{\sf W}(x) &:=& \frac{1}{\nu}\left(2\frac{\nabla\otimes \nabla V}{\nu^2 m^2}\nabla V +\frac{\Delta V \nabla V}{\nu^2 m^2}
+\frac{\beta\hbar^2}{12m^2} \nabla \cdotp \nabla\otimes \nabla V\right)(x) \,
,\\[2mm]
{\sf E}\equiv {\sf E}(x)&:=& \frac{\nabla V(x)}{\nu m}\,,\quad \forall\,x\in \R^d.
\end{eqnarray*}
\begin{assumption}
\label{ass:assumptiononV}
$V$ belongs to $H_x^{k+2}$ with $d$-admissible $k$ and it satisfies the following
$$
\exists \; c>0 \quad\hbox{s.t.}~\,{\sf D}(x) \,y \otimes y \:\;\ge\:\; c |y|^2\,,\quad \forall\,x,y\in \R^d\,,
$$
\end{assumption}
This implies that ${\cal D}$ is a uniformly elliptic differential operator. Thus we can state the following:
\begin{proposition}   \label{prop:existence}
Let $V$ satisfy Assumption \ref{ass:assumptiononV} 
and, in addition, $\nabla\Delta V\in W^{j-1,\infty}_x$
with $j\in \N$. Then the unique global solution $n=n(t)$ of Eq.~\eqref{eq:qdd1} with $n(0)\in L^2_x$ satisfies $n(t)\in H_x^{j}$ for $t>0$, and the following estimate
\begin{equation}
\label{eq:ana}
\| n(t) \|_{H_x^{j}} \leq M_j (\ze t)^{-j/2}\| n(0) \|_{L^2_x} 
\end{equation}
holds with $M_j>0,$ for $\ze,t\to 0^+\,.$
\end{proposition}
In the following, by $\nabla F\in L^2(\R^d_x)$ we mean $\nabla F\in (L^2_x)^d$. Moreover, we consider $0<\ze<1$ and the constants are independent of $\ze$, unless specified.\\

\noindent {\bf Proof}.
By Assumption \ref{ass:assumptiononV}
on the potential $V$, the operator ${\cal D}$ defined on $D({\cal D})=H^2_x$ generates an analytic contraction semigroup $(T(t))_{t\geq 0}$ on $L^2_x$ (cf.~Thm.VI.5.22 of \cite{EngelNagel}).\\
Let us derive here a basic estimate we shall use intensively in the following. By Assumption \ref{ass:assumptiononV}, for all $u\in H^2_x$
$$
\|\nabla\cdotp\nabla u\|_{L^2_x} \leq C \|{\sf D}\nabla\otimes\nabla u\|_{L^2_x}\leq \frac{C}{\ze} (\|{\cal D}u\|_{L^2_x} + \|\mathrm{div}{\sf D}\cdotp \nabla u\|_{L^2_x})\,,
$$
where the second inequality follows from the definition of the operator ${\cal D}$. Moreover, by using that
for all $u\in H^2_x$, 
\begin{equation}
\label{eq:GN}
\|\nabla u\|_{L^2_x}\leq C_{\delta}\|u\|_{L^2_x} + \delta \|\nabla \cdotp\nabla u\|_{L^2_x}\,,\quad \forall\,\delta>0,
\end{equation}
with $C_{\delta}>0$, it holds
$$
\|\mathrm{div}{\sf D}\cdotp \nabla u\|_{L^2_x}\leq \|\mathrm{div}{\sf D}\|_{L^{\infty}_x}(C_{\delta}\|u\|_{L^2_x} + \delta \|\nabla \cdotp\nabla u\|_{L^2_x})\,,\quad \forall\,\delta>0,
$$
with $C_{\delta}>0$. Thus,
$$
\|\nabla\cdotp\nabla u\|_{L^2_x} \leq \frac{c}{\ze(1-\delta \|\mathrm{div}{\sf D}\|_{L^{\infty}_x}/\ze)} (\|\mathrm{div}{\sf D}\|_{L^{\infty}_x}C_{\delta}\|u\|_{L^2_x}+ \|{\cal D}u\|_{L^2_x})\,,
$$
and, in conclusion, for an appropriate choice of $\delta>0$, exists some constant $C>0$ such that
\begin{equation}
\label{eq:stimabase}
\|\nabla\cdotp\nabla u\|_{L^2_x} \leq \frac{C}{\ze} (\|u\|_{L^2_x}+ \|{\cal D}u\|_{L^2_x})\,.
\end{equation}
The operator ${\cal G}$ can be written as ${\cal G}={\cal G}_1+{\cal G}_2$ with ${\cal G}_1 f:=\ze {\sf W} \cdotp \nabla f$ defined on $H^1_x$,
and ${\cal G}_2 f:=\ze \nabla\cdotp {\sf W} f$, defined on $L^2_x$. The operator ${\cal G}_1$ is ${\cal D}$-bounded, i.e., for all $n\in D({\cal D})$,
\begin{eqnarray}
\label{eq:G1_prima}
\|{\cal G}_1 n\|_{L^2_x}&\leq&\ze\| {\sf W} \|_{L^{\infty}_x}C (\|n\|_{L^2_x}+\|\nabla\cdotp\nabla n\|_{L^2_x})\\
&\leq& \|{\sf W}\|_{L^{\infty}_x} (C\|n\|_{L^2_x} +C{\|{\cal D}n\|_{L^2_x}})\nonumber\\
\label{eq:G1_terza}
&\leq& b\|n\|_{L^2_x} + a{\|{\cal D}n\|_{L^2_x}}\,,
\end{eqnarray}
by using \eqref{eq:stimabase} and $\ze<1$. Moreover, the ${\cal D}$-bound $a_1$ defined by
$$
a_1:=\mathrm{inf}\,\{a\geq0\,|\, \exists \,b>0\;\hbox{s.t.~\eqref{eq:G1_terza} holds}\}
$$
is zero, by substituting \eqref{eq:G1_prima} with \eqref{eq:GN}.
The operator ${\cal G}_2 f:=\ze \nabla\cdotp {\sf W} f$, defined on $L^2_x$, is bounded.
The operator ${\cal E}$ can be written as ${\cal E}={\cal E}_1+{\cal E}_2$ with ${\cal E}_1 f:=E \cdotp \nabla f$, defined on $H^1_x$, ${\cal D}$-bounded with ${\cal D}$-bound $a_2=0$, since, for all $n\in D({\cal D})$,
\begin{eqnarray}
\|{\cal E}_1 n\|_{L^2_x}&\leq&\| E\|_{L^{\infty}_x}(C_{\delta}\|n\|_{L^2_x}+\delta\|\nabla\cdotp\nabla n\|_{L^2_x})\nonumber\\
&\leq& C\, \|E\|_{L^{\infty}_x}\left({C_{\delta}}\|n\|_{L^2_x}+\frac{\delta}{\ze} (\|n\|_{L^2_x}+\|{\cal D}n\|_{L^2_x})\right)\,,\quad\forall\, \delta>0\,,\label{eq:E1}
\end{eqnarray}
by using \eqref{eq:GN} and \eqref{eq:stimabase}. The operator ${\cal E}_2 f:= \nabla\cdotp E f$ is  defined on $L^2_x$ and bounded.\\
Thus, by Thm.~III.2.10 of \cite{EngelNagel}, $({\cal D}+{\cal G}+{\cal E},D({\cal D}))$ generates an analytic semigroup on $L^2_x$
that we shall indicate with $(S(t))_{t\geq 0}$. More precisely, it holds
\begin{equation}
\label{eq:stimaanaliticity}
\|({\cal D}+{\cal G}+{\cal E})^{\alpha}S(t)u\|_{L^2_x}\leq M_{\alpha} t^{-\alpha}\|u\|_{L^2_x}, \quad t\to 0^+\,,\;\forall\,\alpha \geq 0
\end{equation}
with $M_{\alpha}$ independent of $\ze$, by employing Lemma III.2.6 of \cite{EngelNagel}.\\
In order to derive estimate \eqref{eq:ana}, let us start from the following inequality
\begin{equation}
\label{eq:stimastandard}
\ze^{m/2}\|u\|_{H^m_x}\leq C\|({\cal D}+{\cal G}+{\cal E})^{m/2}u\|_{L^2_x}
\end{equation}
that is yielded by similar arguments to \eqref{eq:stimabase}.
By combining \eqref{eq:stimastandard} with \eqref{eq:stimaanaliticity}, we get
\begin{equation}
\label{eq:secondastima}
\|S(t) u\|_{H^j_x}\leq C\ze^{-j/2} \|({\cal D}+{\cal G}+{\cal E})^{j/2}S(t)u\|_{L^2_x}\leq C_j (\ze t)^{-j/2}\|u\|_{L^2_x}\,,
\end{equation}
which holds for all $u\in L^2_x$ and for small $t$.
\qed\\
In estimate \eqref{eq:ana} can be easily removed the singularity with respect to $t$:
\begin{corollary}
Let the assumptions of the Prop.~\eqref{prop:existence} hold. In addition, let $n(0)$ belong to $H^j_x$. Then the solution $n(t)$ belongs to $H^j_x$ for all $t>0$ and satisfies
\begin{equation}
\label{eq:primastimabis}
\|n(t)\|_{H^j_x}\leq C\ze^{-j/2} \|n(0)\|_{H^j_x}\,,
\end{equation}
for $\ze,t\to 0^+\,.$
\end{corollary}
{\bf Proof.}\\
The following inequality holds
\begin{equation}
\label{eq:stimastandardbis}
\ze^{m/2}\|u\|_{H^m_x}\leq C\|({\cal D}+{\cal G}+{\cal E})^{m/2}u\|_{L^2_x}\leq C \|u\|_{H^m_x}
\end{equation}
for all $u\in D(({\cal D}+{\cal G}+{\cal E})^{m/2}),$ with $m\leq j$, and for $\ze\to 0^+$ (cf.~\eqref{eq:stimastandard}). Then, in particular,
\begin{multline}
\label{eq:primastima}
\|S(t) u\|_{H^m_x}\leq C\ze^{-m/2} \|({\cal D}+{\cal G}+{\cal E})^{m/2}S(t)u\|_{L^2_x}\\
\leq C\ze^{-m/2}\|({\cal D}+{\cal G}+{\cal E})^{m/2}u\|_{L^2_x}\leq C\ze^{-m/2}\|u\|_{H^m_x}\,,
\end{multline}
which holds for all $u\in D(({\cal D}+{\cal G}+{\cal E})^{m/2}),$ for small $t$ and $\ze$: the first inequality sign corresponds to the first one in Eq.~\eqref{eq:stimastandardbis}, the second inequality follows by exchanging $({\cal D}+{\cal G}+{\cal E})^{m/2}$ with $S(t)$ and the third one comes from Eq.~\eqref{eq:stimastandardbis}.
\qed\\
\begin{remark}
\em
Observe that in the low-field case, the QDD equation looks like Eq.~\eqref{eq:qdd1} with 
$$
{\sf W}(x)=\left(\frac{\beta\hbar^2}{12\nu m^2} \nabla \cdotp \nabla\otimes \nabla V\right)(x)\,,\quad{\sf D}(x)= \frac{1}{\nu}\left(\frac{{\mathcal I}}{\beta m}+\frac{\beta\hbar^2}{12m^2}\nabla\otimes\nabla V\right)(x)\,.
$$
In order to establish well-posedness result and estimate \eqref{eq:ana} for all $j\in \N$, the same assumptions of Prop.~\ref{prop:existence} are required on the potential $V$ and on the initial datum $n(0)$. This result is to be compared with the analysis in \cite{juengelpinnau1}, where it is tackled the fourth-order, non-linear equation obtained by the approximation $\nabla \log n= -\beta \nabla V + {\cal O}(\hbar^2)$, cf.~\cite{Gard94}.
\qedrem\\
\end{remark}
\noindent By increasing the assumptions on the initial datum, we can remove the singular behaviour of the estimate~\eqref{eq:ana} with respect to $t$ and $\ze$.
\begin{corollary}
\label{cor:regularity}
Let V satisfy Assumption \ref{ass:assumptiononV} and $\nabla\Delta V\in W_x^{2j-1,\infty}$. Then the solution $n(t)$ of Eq.~\eqref{eq:qdd1} with $n(0)\in D({\cal D}^j)$ satisfies,  for $\ze, t \rightarrow 0^{+}$
\begin{equation}
\label{stiman1}
\| n(t) \|_{H^j_x} \leq C\| n(0) \|_{H^{2j}_x} \,.
\end{equation}
Moreover, the following refinement holds
\begin{equation}
\label{stiman2}
\| n(t) \|_{H^j_x} \leq C \| n(0) \|_{H^j_x}\,.
\end{equation}
\end{corollary}
{\bf Proof}.
We prove the thesis in the case $j=1$. For $j>1$ the thesis follows by an induction procedure similarly to \cite{banasiakAAM}. Due to the regularity with respect to the variable $x$ of the solution $n(t),$ for $t>0$, we can find the evolution equation for $\nabla n$ by differentiating
$$
\nabla\left(\frac{\partial}{\partial t} n\right)=\frac{\partial}{\partial t}(\nabla n)
= \nabla({\cal D}+{\cal G}+{\cal E})n=({\cal D}+{\cal G}+{\cal E})\nabla n - [({\cal D}+{\cal G}+{\cal E}),\nabla]n,
$$
where we indicate with $[({\cal D}+{\cal G}+{\cal E}),\nabla]$ the commutator among the two operators. Since
$$
-[({\cal D}+{\cal G}+{\cal E}),\partial_k] = \ze\sum_{i,j} \partial_i \left( \partial_k{\sf D}_{ij} \partial_j n \right) +  \sum_i \partial_i\left( \partial_k\left(\ze {\sf W}_i +{\sf E}_i\right) n \right)=:({\cal D}^\prime+{\cal G}^\prime+{\cal E}^\prime) n,
$$
$\nabla n$ satisfies
\begin{equation}
\label{eq:diffnabla}
\frac{\partial}{\partial t}(\nabla n)\:\;=\:\;({\cal D}+{\cal G}+{\cal E})\nabla n +({\cal D}^\prime+{\cal G}^\prime+{\cal E}^\prime) n\,.
\end{equation}
The solution of the previous equation can be expressed by the Duhamel formula via the analytic semigroup $S(t)$ generated by $({\cal D}+{\cal G}+{\cal E})$, as
\begin{equation}
\label{eq:duh1}
\nabla n (t) = S(t) \nabla n (0) + \int_0^t \!\! S(t-s) ({\cal D}^\prime+{\cal G}^\prime+{\cal E}^\prime) n(s)\, ds\,.
\end{equation}
Moreover we can estimate
\begin{eqnarray}
\|\nabla n (t) \|_{L^2_x}&\leq& C\|\nabla n (0)\|_{L^2_x} + C\int_0^t \!\! \| ({\cal D}^\prime+{\cal G}^\prime+{\cal E}^\prime) n(s)\|_{L^2_x}\, ds\nonumber\\
\label{eq:stimarefined}
&\leq& C\|\nabla n(0)\|_{L^2_x} + C \int_0^t \!\!(\ze \|n(s)\|_{H^2_x}+\|n(s)\|_{H^1_x})\,ds\\
&\leq& C \|n(0)\|_{H^2_x}+ C \int_0^t \!\!\|\nabla n(s)\|_{L^2_x}\,ds\,.\nonumber
\end{eqnarray}
by using \eqref{eq:primastimabis} with $j=2$, provided $n(0)\in H^2_x$. Therefore, by Gronwall lemma, we derive \eqref{stiman1}.
In order to prove \eqref{stiman2}, we apply for the function $n(t)=S(t)n(0)$ the first inequality in \eqref{eq:secondastima} with $j=2$ and we obtain
$$
\|n(t)\|_{H^2_x}\leq\frac{C}{\ze}\|({\cal D}+{\cal G}+{\cal E})S(t)n(0)\|_{L^2_x}\,.
$$
Then we use \eqref{eq:stimaanaliticity} for the term  $({\cal D}+{\cal G}+{\cal E})^{1/2}S(t)\left[({\cal D}+{\cal G}+{\cal E})^{1/2}n(0)\right]$ with $j=1$, and we get
\begin{eqnarray}
\|n(t)\|_{H^2_x}&\leq& \frac{C}{\ze}\|({\cal D}+{\cal G}+{\cal E})^{1/2}S(t)({\cal D}+{\cal G}+{\cal E})^{1/2}n(0)\|_{L^2_x}\nonumber\\
\label{eq:stiman1c}
&\leq& \frac{C}{\ze} t^{-1/2}\|({\cal D}+{\cal G}+{\cal E})^{1/2}n(0)\|_{L^2_x}\nonumber\\
&\leq& \frac{C}{\ze} t^{-1/2}\|n(0)\|_{H^1_x}\,,
\end{eqnarray}
where for the last inequality the estimate \eqref{eq:stimastandardbis} with $m=1$ is used. Hence it holds for all $n(0)\in {\cal D}(({\cal D}+{\cal G}+{\cal E})^{1/2})\equiv {H^1_x}$. By using \eqref{eq:stiman1c} in \eqref{eq:stimarefined}, we get \eqref{stiman2} in the case $j=1$.
\qed \\

\noindent
The other (non-null) term of the bulk part is $\psibar_1$, which is of first order in $\ze$. Since it satisfies
$$
\psibar_1 = -({\cal Q}({\cal A}+{\cal C}) \calq)^{-1} \qsp (n M)
$$
(cf.~Eq.~\eqref{eq:psibar1}), by using the definitions \eqref{eq:QSP} and \eqref{eq:qacq}, it can be written explicitly as
\begin{multline}
\label{eq:psi1explicit}
\psibar_1 (x,v,t)= \nabla n (x,t) \cdotp\, { \cal F}^{-1}\left\{\frac{1}{i\delta V-\nu}{\cal F}\left( v M + \frac{M \nabla V}{\nu m}\right)\right\} (x,v)+\\
 n (x,t)\, { \cal F}^{-1}\left\{\frac{1}{i\delta V-\nu}{\cal F}\left( v \cdotp \nabla_xM+ \frac{M \Delta V}{\nu m}\right)\right\} (x,v)\,, \,\forall \,(x,v,t)\in \R^{2d}\times \R^+\,.
\end{multline}
Thus, the estimates for the solution $n$ in the previous corollary are the crucial ingredient to establish well-posedness of the definition \eqref{eq:psi1explicit} and the behaviour with respect to time of the function $\psibar_1$ and of its derivatives.\\
Another fundamental aspect is the shape of the initial datum $n(0)$ for Eq.~\eqref{eq:qdd1}: by \eqref{cx0b}, it is given by $n(0)=n_0+\ze n_1$, and the following estimate holds
\begin{equation}
\label{eq:stimacx0}
\|n(0)\|_{H^j_x}\leq \|n_0\|_{H_x^j}+\ze\|n_1\|_{H_x^{j}}\leq \|w_0\|_{H_k^j}+\ze\, C(\|V\|_{H^{k+j+2}_x})\|w_0\|_{H_{k+1}^{j+1}},
\end{equation}
for all $d$-admissible $k$, by using the estimate \eqref{eq:normQACQH}.
\begin{proposition}   \label{prop:estimates_bulk}
Let $n$ be a solution of the drift-diffusion \eqref{eq:qdd1} with initial value $n(0)$ given by \eqref{cx0b},
with $w_0\in H^4_{k+1}$, and with $V$ satisfying Assumption \ref{ass:assumptiononV} and $\nabla\Delta V\in W_x^{5, \infty}$.
Then $\psibar_1$ is strongly differentiable with respect to $t>0$, and for every $t>0$ it satisfies
$$
\psibar_1(t)\in {\cal D}({\cal Q}({\cal A}+{\cal C}){\cal Q})\cap {\cal D}({\cal Q}{\cal S}{\cal Q})\,.
$$
Moreover there exists a constant $M>0$ such that, for $\ze, t \to 0^+$,
\begin{eqnarray}
\label{stimapsi1diff0}
\left\| \partial_t \psibar_1 (t) \right\|_{X_{k}}&\leq& M \| w_0 \|_{H^4_{k+1}} \,,\\
\label{stimapsi1diff}
\left\| \partial_t \psibar_1 (t) \right\|_{H^1_{k}}&\leq& M (1+1/t)\| w_0 \|_{H^4_{k+1}} \,,\\
\label{stimapsi1Q}
\left\| SQ \psibar_1 (t) \right\|_{H^1_k} &\leq& M \| w_0 \|_{H^4_{k+1}}\,. \end{eqnarray} \end{proposition}
{\bf Proof}.
If we differentiate with respect to $t$ the expression
\eqref{eq:psi1explicit}, the only $t$-dependent functions are $\nabla n$ and $n$, explicitly
\begin{equation}
\label{eq:psi1explicitdiff}
\partial_t \psibar_1 (x,v,t)= \partial_t (\nabla n) (x,t) \cdotp\, A(x,v)+
 \partial_t n (x,t)\,B (x,v)\,, \quad \forall \,(x,v,t)\in \R^{2d}\times \R^+
\end{equation}
where the functions $A_i, B$, defined in \eqref{eq:psi1explicit}, are sufficiently regular, because of the assumptions on $V$. The differentiability of $n$ with respect to $t$ depends on the analiticity of the semigroup $(S(t))_{t\geq 0}$. The differentiability of $\nabla n$, instead, follows from the expression \eqref{eq:duh1}: since each term is continuously differentiable in time, also $\nabla n$ is.\\
Moreover, by using $\partial_t \nabla n=\nabla\partial_t n$ and the evolution equation for $n$,
\begin{eqnarray}
\label{perstimapsi1diff0}
\|{\partial_t}(\nabla n)\|_{L^2_x} &=&\|\nabla({\cal D}+{\cal G}+{\cal E})S( t)n(0)\|_{L^2_x}\nonumber\\
&\leq&\|({\cal D}+{\cal G}+{\cal E})S( t)n_0\|_{H^1_x}+\ze\|({\cal D}+{\cal G}+{\cal E})S( t)n_1\|_{H^1_x}\nonumber\\
&\leq&\|S( t)n_0\|_{H^3_x}+\ze \|S( t)n_1\|_{H^3_x}
\:\;\leq\:\; C (\|w_0\|_{H^3_{k}}+\ze\|w_0\|_{H^4_{k+1}})\nonumber\\
&\leq& C \|w_0\|_{H^4_{k+1}}
\end{eqnarray}
where we split $n(0)=n_0+\ze n_1$ and we use \eqref{stiman2}, together with \eqref{eq:stimacx0}. Similarly,
\begin{eqnarray}
\label{perstimapsi1diff}
\|{\partial_t}(\nabla n)\|_{H^1_x} &\leq& C \left (\|\nabla({\cal D}+{\cal G}+{\cal E})S( t)n(0)\|_{L^2_x}+ \|\nabla\cdotp \nabla({\cal D}+{\cal G}+{\cal E})S( t)n(0)\|_{L^2_x}\right )\nonumber\\
&\leq& C \left(\|w_0\|_{H^4_{k+1}}+ \|({\cal D}+{\cal G}+{\cal E})n_0\|_{H^2_x}+ \frac{\ze}{\ze t}\|({\cal D}+{\cal G}+{\cal E})n_1\|_{L^2_x}\right)\nonumber\\
&\leq& C \left(\|w_0\|_{H^4_{k+1}}+ \|w_0\|_{H^4_{k}}+\frac{1}{t}\|w_0\|_{H^3_{k+1}}\right)\,,
\end{eqnarray}
where the first addendum in the inequality comes from \eqref{perstimapsi1diff0}. The second and the third terms come by exchanging $S(t)$ with $({\cal D}+{\cal G}+{\cal E})$ and using $n(0)=n_0+\ze n_1$, then we apply estimate \eqref{stiman2} to get the second term, and estimate \eqref{eq:ana} to obtain the third term.
Finally, inequality \eqref{perstimapsi1diff} follows from estimates \eqref{eq:stimastandardbis} and \eqref{eq:stimacx0}. In order to prove \eqref{stimapsi1Q}, let us consider again the abstract definition of $\psibar_1$ (see \eqref{eq:psibar1}):
$$
\psibar_1(t)=-({\cal Q}({\cal A}+{\cal C}){\cal Q})^{-1}({\cal Q}S{\cal P})(n(t)M)\,.
$$
Since $\qsp \fibar$ reads (see \eqref{eq:QSP})
$$
\qsp nM = n\left[-v\cdotp \nabla_x  M+M\int\!\! v \cdotp \nabla_xM dv \right]
+\nabla_x n\cdotp
\left[ M\left(-v+ \int\!\! v M dv \right) \right],
$$
under the present hypotheses, $\qsp (nM)$ belongs to $H^2_{k}$, thus $\psibar_1(t)\in {\cal D}({\cal Q}({\cal A}+{\cal C}){\cal Q})\cap {\cal D}({\cal Q}S{\cal Q})$. By
\eqref{eq:psi1explicit},
\begin{multline}
S{\cal Q}\psibar_1(x,v,t)=
\nabla \cdotp \nabla n (x,t)\,  v\,\cdotp A(x,v) \\
+\nabla n (x,t) \cdotp ( v \,\cdotp \nabla \cdotp A+ v B)(x,v) + n(x,t)\, v\,\cdotp \nabla B(x,v)\,.\nonumber
\end{multline}
Thus, in order to estimate $\|S{\cal Q}\psibar_1(t)\|_{X_k}$ and $\|S{\cal Q}\psibar_1(t)\|_{H^1_k}$, it is necessary to
evaluate $\|\nabla \cdotp \nabla n (t)\|_{L^2_x}$ and $\|\nabla \cdotp \nabla n (t)\|_{H^1_x}$, $\|\nabla n(t)\|_{L^2_x}$
and $\|\nabla n(t)\|_{H^1_x},$ respectively.
In particular,
$$
\|\nabla\cdotp\nabla n(t)\|_{H^1_x}\leq C\| S(t)n(0)\|_{H^3_x}\leq C\|n(0)\|_{H^3_x}\leq C \|w_0\|_{H^3_k}+\ze C\|w_0\|_{H^4_{k+1}}\,,
$$
again by \eqref{stiman2}. Thus, we can conclude by using the regularity properties of $A_i, B$.
\qed\\
\begin{remark}
{\em Observe that it is possible to remove the singularity for $t\to 0^+$ in the estimate \eqref{stimapsi1diff}, by assuming $w_0\in H^5_{k+1}$ and modifying last two lines of \eqref{perstimapsi1diff} as follows
\begin{eqnarray}
\label{perstimapsi1diff1}
\|{\partial_t}(\nabla n)\|_{H^1_x}
&\leq& C \left(\|w_0\|_{H^4_{k+1}}+ \|({\cal D}+{\cal G}+{\cal E})n_0\|_{H^2_x}+ {\ze}\|({\cal D}+{\cal G}+{\cal E})n_1\|_{H^2_x}\right)\nonumber\\
&\leq& C \left(\|w_0\|_{H^4_{k+1}}+ \|w_0\|_{H^4_{k}}+ \ze \|w_0\|_{H^5_{k+1}}\right)\,.
\end{eqnarray}
}
\qedrem\\
\end{remark}
\section{Estimate of the error}
\setcounter{equation}{0}
In this section we prove rigorously that the high-field QDD equation, originated by the asymptotic
expansion up to the first order in $\ze$, is an approximation of order $\ze^2$ of the high-field
Wigner-BGK system \eqref{system}. To this aim, we consider the errors obtained by replacing the
functions ${\cal P}w=\zf$ and ${\cal Q}w=\psi$ by the terms of their expansion up to first
order in $\ze$.
We shall prove the following
\begin{theorem}
\label{maintheorem}
If the initial value $w_0$ belongs to ${H^4_{k+2}}$ and $V$ satisfies
Assumption \ref{ass:assumptiononV} and $\nabla \Delta V \in W_x^{5,\infty}$,
then for any $T$, $0<T<\infty$, there is a constant $C$ independent of
$\ze$ such that 
\begin{equation}
\label{eq:finalestimate}
\left\|\zf(t) + \psi(t)- [\fibar(t) + \ze \psibar_1(t) + \psitil_0(t/\ze) +\ze \fitil_1(t/\ze)+ \ze \psitil_1(t/\ze)] \right\|_{X_k} \leq C \ze^2 \,,
\end{equation}
uniformly for $ 0 \le t \le T$.
\end{theorem}
This result relies on the estimates established in Propositions \ref{prop:estimates_initial}, \ref{prop:estimates_bulk},
about the behaviour with respect to time of the initial layer functions
$\fitil_1$, $\psitil_1$ and the bulk functions.
Let us split the error in two contributions
\begin{equation}
\label{eq:def_errors}
  y(t)  
  =\zf(t)  - [\fibar(t) + \ze \fitil_1(\zt)]\,,\qquad
  z(t)=
  \psi(t) - [\psitil_0(\tau)+\ze \psibar_1(t) + \ze \psitil_1(\zt) ]
\end{equation}
where $ \zt = \frac{t}{\ze}$.
The evolution equations for the errors $y$ and $z$ can be deduced from those satisfied by their components (cf.~systems \eqref{sispro},\eqref{sisprobis},\eqref{initial}). Hence, we have
\begin{equation}
\left\{
\begin{array}{lcl}
 \displaystyle \frac{\pt y}{\pt t} & = & \psp y + \psq z + f \\[4mm]
  \displaystyle  \frac{\pt z}{\pt t} & = & \qsp y + \qsq z +
                 \displaystyle \frac{1}{\ze} {\cal Q(A+C)Q} z + g
\end{array} \right.                                          \label{errorsystem}
\end{equation}
with initial conditions
$$
y(0) = 0\,, \qquad z(0) = 0 \,,
$$
and inhomogeneous terms $f$ and $g$ defined by \begin{eqnarray*}
f(t) &=&  \ze \,\left[\psp \fitil_1(\zt) + \psq \psitil_1(\zt)\right]  \\
g(t) &=& \ze \left[- \frac{\pt \psibar_1}{\pt t} + \qsq \psibar_1(t) + \qsp \fitil_1(\zt)
                   + \qsq \psitil_1(\zt)  \right]\,.
\end{eqnarray*}
It is convenient to separate the evolution of the error relative to initial layer part from the one corresponding to the bulk part. Let us define
$$ r = y + z = r_i + r_b $$
with
\begin{equation}
\label{rirb} r_i = - \ze \fitil_1
- \psitil_0 - \ze \psitil_1 \,,\qquad
 r_b
= \zf + \psi - \fibar - \ze \psibar_1 \,.
\end{equation}
The derivation of estimate \eqref{eq:finalestimate} is split according to \eqref{rirb} in the next two Lemmata.
\begin{lemma}
Under the assumptions $V\in H^{k+4}_x$ and $w_0\in H^2_{k+2}$, for any $T$, $0<T<\infty$, there is a constant $C$ independent of
$\ze$  such that
\begin{equation}
\label{eq:finalestimatei}
\| r_i(t) \|_{X_k} \leq C \ze^2 \,,
\end{equation}
uniformly for $ 0 \le t \le T$.
\end{lemma}
{\bf Proof}.
The initial layer error $r_i= r_i(t)$ satisfies the equation
\begin{equation}\label{eqri}
 \frac{\pt r_i}{\pt t} (t)= S r_i(t) + \frac{1}{\ze} {\cal (A+C)} r_i(t) + \ze S(\fitil_1 + \psitil_1) \left(\frac{t}{\ze}\right) \,, \quad r_i(0)=0\,.
 \end{equation}
The operator $S + {\cal (A+C)}/\ze$ generates an uniformly bounded semigroup in $X_k$, $Z(t)$, cf.~\cite{FroPavVan}. Thus, 
the mild solution of \eqref{eqri} is given by
$$
r_i(t) =
Z(t) r_i(0) + \ze \int_0^t Z(t-s) S(\fitil_1 + \psitil_1) (s/\ze) ds \,,
$$
with
\begin{eqnarray}
\|r_i(t)\|_{X_k} &\leq& C \ze  \int_0^t 
\|S(\fitil_1 + \psitil_1)\left({s}/{\ze}\right)\|_{X_k} \, ds \,.
\label{mildri}
\end{eqnarray}
The estimate of $\|S(\fitil_1 + \psitil_1)\left({s}/{\ze}\right)\|_{X_k}$ is a bit tedious, thus we simply sketch it. It is convenient to use the projections ${\cal P}, {\cal Q}$ and evaluate $\psp \fitil_1$, $\qsp \fitil_1$, $\psq\psitil_1$, and $\qsq\psitil_1$ separately. By their definitions (cf.~\eqref{eq:PSP},\eqref{eq:QSP}) $\psp,\qsp\in {\cal L}(H^1_{k},X_k)$, provided $V\in H^{k+2}_x$, thus, it holds the following modification of the estimate \eqref{eq:tilde_zf_1proof} for $\fitil_1$,
 \begin{eqnarray}
\label{eq:tilde_zf_1proofbis}
 \|\psp\tilde\zf _1(\tau)\|_{X_k} &\leq & \||\psp\|||\|\psq[{\cal Q(A+C)Q}]^{-1}\|| \; \|G(\tau)\psi_0\|_{H^2_{k+1}}
\nonumber\\
&\leq&{\rm M}\,e^{-\nu_{k+3}\tau}\|\psi_0\|_{H^2_{k+1}}\,,
\end{eqnarray}
since $ \psq[{\cal Q(A+C)Q}]^{-1}\in{\cal L}(H_{k+1}^2,H^1_k)$, provided $V\in H^{k+4}_x$.
Analogously,
 \begin{eqnarray}
\label{eq:tilde_zf_1prooftris}
 \|\qsp\tilde\zf _1(\tau)\|_{X_k} &\leq & \||\qsp\|||\|\psq[{\cal Q(A+C)Q}]^{-1}\|| \; \|G(\tau)\psi_0\|_{H^2_{k+1}}
\nonumber\\
&\leq&{\rm M}\,e^{-\nu_{k+3}\tau}\|\psi_0\|_{H^2_{k+1}}\,.
\end{eqnarray}
Let us recall the expression for $\tilde\psi_1$ (cf.~\eqref{eq:tildepsidef})
$$
\tilde\psi_1(\tau) = G(\tau)\tilde \psi_1(0) + \int_{0}^{\tau}G(\tau -
\sigma)\qsq G(\sigma)\psi_0d\sigma\,.
$$
We evaluate $\| \psq \psitil_1 (\tau) \|_{X_k}$ and $ \| \qsq \psitil_1 (\tau) \|_{X_k}$. Both $\psq$ and $\qsq$ belong to ${\cal L}(H^1_{k+1},X_k)$. Moreover $[{\cal Q(A+C)Q}]^{-1}\qsp\in {\cal L}(H^1_k, X_k)$ by definition, provided $V\in H^k_x$; thus
\begin{eqnarray*}
\|\qsq G(\tau)\tilde\psi_1(0)\|_{X_k}
&\leq& |\|\qsq G(\tau)\||\|({\cal Q(A+C)Q})^{-1}\qsp\zf_0\|_{H^1_{k+1}}\\
&\leq& K{\mathrm e}^{-\nu_k \tau}\|\zf_0\|_{H^2_{k+1}}\,,
\end{eqnarray*}
provided $V\in H^{k+2}_x\,.$ Concerning the second term, we obtain
\begin{eqnarray*}
\left\|\qsq\int_{0}^{\tau}G(\tau -
\sigma)\qsq G(\sigma)\psi_0d\sigma\right\|_{X_k} 
\!\!\!\!\!\!\!\!&\leq& |\|\qsq \||\left\|\int_{0}^{\tau}G(\tau -
\sigma)\qsq G(\sigma)\psi_0d\sigma\right\|_{H^1_{k+1}}\\
\!\!\!\!\!\!\!\!&\leq&K{\mathrm e}^{-\nu_{k+2} \tau}\int_{0}^{\tau}{\mathrm e}^{(\nu_{k+2}-\nu_{k+4}) \sigma}\|\zf_0\|_{H^2_{k+2}}\\
\!\!\!\!\!\!\!\!&\leq& K{\mathrm e}^{-\nu_{k+4} \tau}\|\zf_0\|_{H^2_{k+2}}\,,
\end{eqnarray*}
provided $V\in H_x^{k+4}$.
In conclusion,
$$
\|\qsq \psitil_1 (\tau)\|_{X_k}\leq K{\mathrm e}^{-\nu_{k+4} \tau}\|\zf_0\|_{H^2_{k+2}}\,,
$$
and analogously,
$$
\|\psq \psitil_1 (\tau)\|_{X_k}\leq L\, {\mathrm e}^{-\nu_{k+4} \tau}\|\zf_0\|_{H^2_{k+2}}\,,
$$
for some constant $L>0$.
Finally, it is possible to find constants $\bar\nu>0$ and $\Mbar(\|V\|_{H_x^{k+4}})>0$ such that
$$
\| S(\fitil_1 + \psitil_1) (\tau) \|_{X_k} \le \Mbar(\|V\|_{H_x^{k+4}}) \,{\rm e}^{-\bar \nu \tau} \| w_0 \|_{H^2_{k+2}}.
$$
Coming back to the estimate \eqref{mildri} of $r_i$, for any time $t$ we have
$$
\|r_i(t)\|_{X_k} = C \Mbar \ze  \int_0^t {\rm e}^{-\bar\nu s/\ze} \| w_0 \|_{H^2_{k+2}}
  ds \leq
C  \| w_0 \|_{H^2_{k+2}} \ze^2.
$$
\qed
\begin{lemma}
Under the same assumptions of Proposition \ref{prop:estimates_bulk}, for any $T$, $0<T<\infty$, there is a constant $C$ independent of
$\ze$ such that
\begin{equation}
\label{eq:finalestimateb}
\| r_b(t) \|_{X_k} \leq C \ze^2 \,,
\end{equation}
uniformly for $ 0 \le t \le T$.
\end{lemma}
{\bf Proof}.
The error of the bulk part of the asymptotic expansion satisfies
\eqref{errorsystem} with $f=0$
and, instead of $g$,
\begin{eqnarray*} g_b(t) &=& \ze \left[- \frac{\pt \psibar_1}{\pt t} + \qsq
\psibar_1(t) \right] \, . \end{eqnarray*}
Since the inhomogeneous term $g_b(t)$ has a non uniform
behaviour with respect to $\ze$ for small times, we split the inhomogeneous term $g(t)$
into the sum of two functions, say $g_{b0}$ and $g_{b1}$, as follows
$$
g_{b0}(t) =  \eta_\ze g_b(t) \,, \qquad g_{b1} =  g_b(t) -  g_{b0}(t)\,,
$$
where $ \eta_\ze$ is a not increasing $C^\infty$-function  such that
$$
 \eta_\ze(t) =
\left\{\!\!\!\!\!\!\!\! \begin{array}{lclcc}
& \, 1 &\quad& {\rm for}  &t<\ze/2  \,,\\[-2mm]
\\
& \, 0 &\quad& {\rm for}  &t > 3\ze/2  \,. \end{array} \right.
$$
We write the error $r_b$ as the sum of two parts $r_b = r_{b0} + r_{b1}$, solving the equation
$$
\label{eqribis} \frac{\pt r_{b0}}{\pt t} = S r_{b0} +
\frac{1}{\ze} {\cal (A+C)} r_{b0} + \ze g_{b0} \,,\;\;\; r_{b0}(0)=0\,,
$$
and an analogous one with the inhomogeneous term $g_{b1}$.
Concerning the error $r_{b0}$, the following estimate holds by using Prop.~\ref{prop:estimates_bulk}
\begin{eqnarray*}
\|r_{b0}(t)\|_{X_k}  &\le&  K \ze \int_0^{3\ze/2} \|g_{b0}(s)\|_{X_k} ds
                     \:\;\le\:\;  K \ze \int_0^{3\ze/2}
\left(
\left\| \frac{\partial \psibar_1}{\pt s} (s) \right\|_{X_k} + \left\| SQ \psibar_1 (s) \right\|_{X_k}
\right) \\
&\le&  K \ze \int_0^{3\ze/2}  \| w_0 \|_{H^4_{k+1}}  ds  \le  K  \| w_0 \|_{H^4_{k+1}} \ze^2\,.
\end{eqnarray*}
Finally, we consider the evolution equation for $r_{b1}$: we decompose again such an error as
$$
r_{b1} = \hat r_{b1}  + h(t)\,,
$$
by introducing the auxiliary function $h$, which solves the problem
$$
\frac{\pt h}{\pt t} = \frac{1}{\ze} {\qacq} h + \ze g_{b1} \,,\;\;\; h(0)=0\,.
$$
Consequently, the function $\hat r_{b1}$ satisfies the initial value problem
$$
 \frac{\pt \hat r_{b1}}{\pt t} = S \hat r_{b1} +
\frac{1}{\ze} {\cal (A+C)} \hat r_{b1} +  S {\cal Q} h \,,\;\;\; \hat r_{b1}(0)=0\,,
$$
thus it can be easily estimated in terms of the auxiliary function $h$ as
$$
\|\hat r_{b1}(t)\|_{X_k}  \le  \int_{\ze/2}^t\| S {\cal Q} h(s)ds \|_{X_k} \,ds\,.
$$
Again by the properties of the operator ${\cal (A+C)}$, the solution reads as follows
$$
 h(t) =
\left\{\!\!\!\!\!\!\!\! \begin{array}{lclcc}
& \, 0 &\quad& {\rm for}  &t<\ze/2  \,,\\[-2mm]
\\
& \, \ze \int_{\ze/2}^t G_{\ze}(t-s) g_{b1}(s) ds
 &\quad& {\rm for}  &t > \ze/2  \,, \end{array} \right.
$$
with $G_{\ze}(\zt)$ bounded semigroup generated by $(1/\ze)\cal Q(A+C)Q$.
\begin{eqnarray*}
\label{rbcaporalebis}
\|\hat r_{b1}(t)\|_{X_k}  &\le&  \ze  \int_{\ze/2}^t\| S {\cal Q} h(s)ds \|_{X_k} ds
\\
&\le&
  \ze K \int_{\ze/2}^t \int_{\ze/2}^s      {\rm e}^{-\nu_{k+1} \frac{s-s'}{\ze}}\|g_{b1}(s') \|_{H^1_k} ds' ds \\
  &\le&
  \ze K \int_{\ze/2}^t   \int_{\ze/2}^s      {\rm e}^{-\nu_{k+1} \frac{s-s'}{\ze}} \left(1+\frac{1}{s} \right) \| w_0 \|_{H^4_{k+1}}ds' ds \\
 &\le&
 K \| w_0 \|_{{H^4_{k+1}}}{\ze^2}\,,
\end{eqnarray*}
by applying again Prop.~\ref{prop:estimates_bulk},
for any $t \in [0,T]$, where the constants $K$  depend on $T$.
In conclusion,
$$
\|r_{b}(t)\|_{X_k}  \le K \| w_0 \|_{{H^4_{k+1}}}{\ze^2}\,.
$$
\vskip 20pt
{\bf Acknowledgements. }
The authors are grateful to Luigi Barletti and Jacek Banasiak, for many helpful discussions on the
position of the problem.
This work was performed under the auspices of the {\it National Group for Mathematical Physics} of the
{\it Istituto Nazionale di Alta Matematica} and was partly supported by the {\it Italian Ministery of
University (MIUR} National Project {``Mathematical Problems of Kinetic Theories", Cofin2004).}

\end{document}